\documentclass{emulateapj} \usepackage{psfig} \usepackage{apjfonts}

\usepackage{amssymb}
\usepackage{amsmath}
\newcommand{\be}{\begin{equation}}
\newcommand{\ee}{\end{equation}}
\newcommand{\rev}{}

\shorttitle{The GRB host luminosity and stellar mass functions} 
\shortauthors{Trenti et al.}

\begin{document}



\title{The luminosity and stellar mass functions of GRB host galaxies:
Insight into the metallicity bias}

\author{Michele Trenti\altaffilmark{1,2,\dag}, Rosalba
  Perna\altaffilmark{3}, Raul Jimenez\altaffilmark{4,5}}

\altaffiltext{1}{Institute of Astronomy and Kavli Institute for
  Cosmology, University of Cambridge, Madingley Road, Cambridge, CB3
  0HA, United Kingdom} 
\altaffiltext{2}{School of Physics, The University of Melbourne, VIC 3010, Australia}
\altaffiltext{3}{Department of Physics and Astronomy,  Stony Brook University, Stony Brook, NY 11794-3800, USA}
\altaffiltext{4}{ICREA \& ICC, University of Barcelona, Marti i Franques 1, 08028 Barcelona, Spain}
\altaffiltext{5}{Institute for Applied Computational Science, Harvard University, MA 02138, USA}
\altaffiltext{\dag}{Kavli Institute Fellow}
\email{mtrenti@unimelb.edu.au}

%

\begin{abstract}
 
  Long-Duration Gamma-Ray Bursts (GRBs) are powerful probes of the
  star formation history of the Universe, but the correlation between
  the two depends on the highly debated presence and strength of a
  metallicity bias. To investigate this correlation, we use a
  phenomenological model that successfully describes star formation
  rates, luminosities and stellar masses of star forming galaxies, and
  apply it to GRB production. {\rev We predict the luminosities,
    stellar masses, and metallicities of host galaxies depending on
    the presence (or absence) of a metallicity bias. Our best fitting
    model includes a moderate metallicity bias, broadly consistent
    with the large majority of the long-duration GRBs in metal-poor
    environments originating from a collapsar (probability $\sim
    83\%$, with $[0.74;0.91]$ range at 90\% confidence level), but
    with a secondary contribution ($\sim 17\%$) from a
    metal-independent production channel, such as binary
    evolution. Because of the mass-metallicity relation of galaxies,
    the maximum likelihood model predicts that the metal-independent
    channel becomes dominant at $z\lesssim 2$, where hosts have higher
    metallicities and collapsars are suppressed. This possibly
    explains why some studies find no clear evidence of a metal-bias
    based on low-$z$ samples. However, while metallicity predictions
    match observations well at high redshift ($z\gtrsim 2$), there is
    tension with low redshift observations, since a significant
    fraction of GRB hosts are predicted to have (near) solar
    metallicity. This is in contrast to observations, unless obscured,
    metal-rich hosts are preferentially missed in current datasets,
    and suggests that lower efficiencies of the metal-independent GRB
    channel might be preferred following a comprehensive fit that
    includes metallicity of GRB hosts from \emph{complete}
    samples. Overall, we are able to clearly establish the presence of
    a metallicity bias for GRB production, but continued
    characterization of GRB host galaxies is needed to quantify its
    strength. Tabulated model predictions are available in electronic format.}

\end{abstract}


\section{Introduction}\label{sec:intro}

Long-duration ($t\gtrsim 2~\mathrm{s}$) Gamma Ray Bursts, simply
indicated as GRBs throughout this paper, are widely considered
powerful tracers of the star formation history of the Universe
\citep{bloom2002,chary2007,savaglio09,robertson12}. In fact, their
progenitors are established to be short-lived, massive stars
\citep{galama1998,hjorth2003,woosley2006,berger2011} and the GRBs and
their afterglows are sufficiently luminous to be detected as early as
$\sim 500~\mathrm{Myr}$ after the Big Bang ($z\sim 9.4$, see
\citealt{salvaterra09,tanvir09,cucchiara11}). Furthermore, the
detectability of GRBs is not limited by how faint their host galaxies
might be, making them potential tracers of the \emph{total} star
formation rate at high-redshift, unlike surveys of Lyman-Break
Galaxies (LBGs), which trace star formation only in systems above the
survey detection limit \citep{trenti2013}. This advantage of GRBs over
LBGs as star formation tracers is highlighted by the non-detection of
the majority of GRB host galaxies at $z\gtrsim 5$, which indicates
that even the deepest observations of LBGs with the Hubble Space
Telescope (HST) are only capturing the tip of the iceberg of star
formation during the first billion years after the Big Bang
\citep{trenti12,tanvir12}.

However, the use of GRBs as star formation tracers presents
significant challenges as well. First, the sample size of GRBs at
high-$z$ is still very small, introducing significant Poisson
noise. For example, there are only $\sim 40$ GRBs at $z\gtrsim 3.5$
and a handful at
$z>6.5$\footnote{\url{http://swift.gsfc.nasa.gov/archive/grb\_table/}},
whereas more than 11000 LBGs have been identified at $z\gtrsim 3.5$,
and more than 800 at $z\gtrsim 6.5$ \citep{bouwens2014}. Second,
several studies suggest that the connection between GRB rate and star
formation rate is likely biased by the metallicity of the GRB
progenitor, with low-metallicity environments having a higher yield of
GRBs.

The presence of a metallicity bias has both observational and
theoretical support: Several groups found that host galaxies of GRBs
appear to lie below the typical mass-metallicity relation observed for
star-forming galaxies \citep{fynbo2003,prochaska2004,modjaz2006,
  fruchter2006,thoene2007,graham2013,wang2014}, and this result fits
well within the leading theoretical framework for GRB engine, the
\emph{collapsar} model \citep{woosley1993,macfadyen1999}. According to
this model, the GRB progenitor is a massive star whose rapidly
rotating core collapses to form a black hole, while the outer envelope
falls back and forms an hyperaccreting disk (if endowed with
sufficient angular momentum), powering an energetic relativistic jet.
Since stellar mass losses grow with increasing metallicity, and carry
away precious angular momentum from the star, an enhancement of the
GRB rate in low-metallicity environments is expected
\citep{woosley2006,yoon06,nuza2007,perna2014}, possibly with a sharp cut-off
around solar metallicity \citep{stanek2006,yoon06}. However, this
scenario for GRB production is at odds with other observational
findings of GRB host galaxies that have solar or even super-solar
metallicity \citep{berger2007,levesque10c,savaglio12}, and alternative
models to the \emph{collapsar} scenario have been proposed, such as
progenitors in binaries \citep{fryer05,cantiello2007}, which are less
influenced by metallicity.

Such a complex mix of observational and theoretical results on the
presence, and strength, of a metallicity bias arguably represents the
most significant limiting factor in the use of GRBs as reliable
tracers of star formation
(\citealt{stanek2006,kewley2007,modjaz2008,jimenez13}, but see the
recent paper by \citealt{hunt2014}). A deeper, and clearer,
understanding of how metallicity influences the relation between GRB
and star formation rate is needed to be able to account and correct
for the bias. In this respect, there are two separate questions to
address:
\begin{itemize}
\item  Are GRBs more probable in low-metallicity environments compared to
higher metallicity ones?
\item Is there a maximum metallicity cut-off? 
\end{itemize}
Both affect the connection between GRB and star formation, with an
impact that in general varies over redshift, reflecting the evolution in
the chemical enrichment of galaxies throughout cosmic history. 

GRB host galaxy studies are an ideal tool to understand if a
metallicity bias is present, and how strong it is. As such, they have
been the subject of extensive studies (we refer the reader to the
comprehensive review by \citealt{levesque2014}), {\rev but no clear
  consensus has been reached in the community. Some investigations
  point toward a clear presence of a metallicity-dependence in the
  production of GRBs (e.g., \citealt{boissier2013,lunnan2014}), while
  others conclude that GRBs appear to be unbiased tracers of star
  formation (e.g., \citealt{fynbo2008,kohn2015}), and some highlight that discrimination
  between scenarios is challenging (e.g.,
  \citealt{pontzen2010,laskar2011}). }

Ideally, spectroscopic samples are the most suited for the task {\rev
  of quantifying the bias}, since they allow to establish a direct
connection between GRBs and host metallicity.  However, measuring the
metallicity of galaxies at high redshift is challenging, and generally
limited only to the brightest objects. A wider study extending to
fainter hosts is possible by characterizing luminosities and stellar
masses of GRB host galaxies from broad-band photometry, and then
comparing sample properties against those of LBGs. Qualitatively, a
GRB preference for low-metallicity environments is reflected into
steeper luminosity and mass functions at the faint end compared to a
scenario without bias, because of the established existence of a
mass/luminosity vs. metallicity relation
\citep{Panter03,Panter08,maiolino08,mannucci2010}.  Similarly, the
presence of a metallicity cutoff would introduce a truncation of the
luminosity/mass function for the brightest and most massive hosts.

Even in absence of a metallicity bias, the luminosity and mass
functions of GRB hosts differ from those of LBGs because the GRB rate
is proportional to the star formation rate (SFR), making the host
skewed toward inclusion of systems with higher SFR (i.e. luminosity),
hence resulting in a shallower luminosity function. If
$\phi_{(LBG)}(L)$ is the luminosity function of LBGs, then under the
basic assumption that the luminosity density and the star formation
rates are proportional and that there is no metallicity bias, it
follows that the GRB host luminosity function is proportional to
$\phi_{(LBG)}(L)\times L$. There is, though, one second order but
subtle and crucial correction that needs to be made to this relation
to properly quantify how any deviation from it is connected to the
metallicity bias. That is proper modeling of dust\footnote{For some
  discussion on the influence of dust on deriving physical properties
  of galaxies from their stellar populations see
  e.g. \citet{vespa}}. In fact, the GRB rate depends on the intrinsic
star formation rate, hence on the intrinsic luminosity, not on the
observed one. As a result, the scaling for the GRB luminosity function
(again with no bias) goes as $\phi_{(GRB)}(L_{obs}) \propto
\phi_{(LBG)}(L_{obs}) \times L_{int}$. Since $L_{int}/L_{obs}$ depends
on the metallicity (higher dust content is typically associated with
higher metallicity environments), this effect may partially mask the
presence of a metallicity bias, unless proper dust treatment is taken
into account in the data-model comparison.

Finally, it is also important to recall that, since all the proposed
GRB engines are generically associated with the death of massive stars,
it is generally expected that studies which focus on luminosity
functions at wavelengths that are good tracers of \emph{recent} star
formation are those most suited for the purpose of understanding the
connection between GRBs and star formation. In this respect, the two
most natural choices are UV or far-IR. In this work, we focus on the
first, motivated primarily by the availability of high quality UV
luminosity functions for LBGs. To present a comprehensive analysis, we
include stellar mass functions in the modeling, but the latter
are not as powerful because the connection between star formation and
stellar mass is not as tight as it is with UV luminosity. For example,
the most massive elliptical galaxies in the local universe have little
to no recent star formation, making them highly unlikely to host
GRBs. 

To complement and augment the previous studies of the metallicity bias
of GRB host galaxies, we take a novel approach, starting from a
successful modeling framework that we developed to investigate how the
luminosity function of LBGs evolves with redshift and how it is
connected to the underlying dark matter halo mass function. In
\citet{trenti10} we constructed a first link between luminosity and
dark matter halo mass functions, quantitatively predicting the LBG
luminosity function during the epoch of reionization. In particular,
the model predicted an accelerated decrease of the luminosity density
of galaxies at $z\gtrsim8$ and a steepening of the faint-end slope of
the luminosity function, with both predictions recently verified
thanks to the array of new Hubble observations of $z\gtrsim 8$
galaxies
\citep{bradley12,schmidt2014,ellis2013,mclure13,bouwens2014}. We also
applied our luminosity function model to interpret the non-detections
of GRB host galaxies at $z>5$ by \citet{tanvir12}, concluding {\rev
  (see \citealt{trenti12})} that the majority of star formation at
$z>6$ is happening in faint systems below the current detection limit
(a conclusion deriving naturally from the luminosity function
steepening; {\rev see also \citealt{salvaterra13} which supports
  earlier results}). More recently, we extended our modeling to
achieve a comprehensive description of the LBG UV luminosity function
evolution from $z\sim 0$ to $z\sim 10$, capturing at the same time the
stellar mass density and specific star formation rate evolution
\citep{tacchella13}. We hence applied the model to investigate the
connection between GRB rate and star formation rate and how it varies
with redshift and with the presence of a metallicity bias
\citep{trenti2013}. Our key conclusions were: (1) the best model for
the GRB rate is one that includes production at low metallicity
primarily through the collapsar engine ($\sim 75-80\%$), with
metallicity bias, combined with a second, metal-independent channel
($\sim 20-25\%$), such as a progenitor star in a binary system; and
(2) the metallicity bias becomes negligible to first approximation at
$z\gtrsim 4$, since the majority of star forming galaxies have low
metallicity $Z\lesssim 0.1 Z_{\odot}$. While the conclusions reached
were interesting and shed some new light on the problem, our study was
limited by the uncertainty in the comoving GRB rate.

In this work, we aim at building upon our previous framework to expand
the data-model comparison to UV luminosity functions, stellar mass
functions, {\rev and metallicities} of GRB hosts, with the goal of
providing a more stringent and rigorous characterization of the
metallicity bias. We extend our previous modeling by taking into
account other effects, such as dust obscuration, that are likely to
influence the data-model comparison. In addition, by presenting
predictions for the redshift evolution of GRB hosts luminosity and
stellar mass function, we make testable predictions that can guide the
design of the next generation of surveys aimed at characterizing their
properties. The paper is organized as follows. Section~\ref{sec:model}
describes our model, whose results are presented in
Section~\ref{sec:results}, and then compared to the current
observations in Section~\ref{sec:tough}. Section~\ref{sec:con}
summarizes and concludes with an outlook for the future.

Throughout the paper we use the latest $\Lambda CDM$ concordance
cosmological model with parameters determined by the \emph{Planck}
mission: $\Omega_{\Lambda,0}=0.685$, $\Omega_{m,0}=0.315$,
$\Omega_{b,0}=0.0462$, $\sigma_8=0.828$, $n_s=0.9585$, $h=0.673$
\citep{planck}.

\section{Modeling: GRB rate and host galaxy properties}\label{sec:model}

Following the framework developed in \citet{trenti2013}, we base our
modeling on linking star formation to the assembly of dark-matter
halos. We assume that each halo converts a fraction of its total mass
$\xi(M_h)$ into stars ($M_*=\xi(M_h) \times M_h$) {\rev with a
  constant star formation rate} over the timescale defined by the
halo-assembly time $t_{1/2}(M_h,z)$, that is the time needed to grow
from $M_h/2$ to $M_h$. $\xi(M_h)$ depends on mass but not on redshift
\citep{tacchella13,behroozi13}, while $t_{1/2}$ decreases both for
increasing mass and redshift (e.g., see \citealt{lacey93}). The UV
luminosity of the galaxies in our model is computed {\rev via Stellar
  Population models}, assuming a Salpeter initial mass function from
$0.1~\mathrm{M_{\sun}}$ to $100~\mathrm{M_{\sun}}$
\citep{bruzual03}. At fixed halo (and stellar) mass, higher $z$ halos
form stars on a shorter timescale, hence they have higher UV
luminosity {\rev (see \citealt{tacchella13,trenti2013} for further
  details)}.

The star formation efficiency $\xi(M_h)$ is calibrated at one
reference redshift, thanks to abundance matching between the galaxy
luminosity function and the dark-matter halo mass function. After such
calibration, the model has no free parameters, and the evolution of
the dark-matter halo mass function and of the halo assembly time fully
determines the star formation rate and galaxy luminosity function at
all other redshifts. As discussed in \citet{tacchella13}, this minimal
model with no free parameters is remarkably successful in capturing
the evolution from $z\sim 0$ to $z\sim 10$ of (1) the UV luminosity
density (star formation rate); (2) the galaxy luminosity function; and
(3) the stellar mass density and specific star formation rate. Like in
our previous work, we calibrate $\xi(M_h)$ at $z=4$. Here, we use the
\citet{sheth99} halo mass function and the latest determination of the
observed UV luminosity function
[$\phi(L)=\phi^*(L/L^*)^{\alpha}\exp{-(L/L^*)}$] derived by
  \citet{bouwens2014}:
  $\phi^*=1.35^{+0.22}_{-0.19}\times10^{-3}~\mathrm{Mpc^{-3}}$;
  $\alpha=-1.65\pm 0.03$ and $M_{AB}^{(*)}=-21.08$, where
  $M^{(*)}=-2.5\log_{10}{L^*}$. We assume that the LF extends to
  fainter than observed luminosity, down to $M_{UV}\leq -11$, which
  corresponds to $M_h\gtrsim 5\times 10^8~\mathrm{M_{\odot}}$ at
  $z\sim 4$ (e.g. see \citealt{trenti10,tacchella13}). While this
  represents a significant extrapolation compared to the typical
  magnitude limit of Lyman-Break galaxy surveys ($M_{UV}\sim -17$, see
  \citealt{bouwens07}), observations of LBGs behind gravitational
  lenses have shown that the LF of LBGs at $z\sim 2$ continues to be a
  steep power law at $M_{UV} \sim -13$ \citep{alavi13}. Note that star
  forming sites with UV magnitude above $M_{AB}>-11$ are unlikely to
  host a GRB in any case, since their star formation rate is so low
  ($\dot{\rho_*} \lesssim 1.3\times
  10^{-3}~\mathrm{M_{\sun}~yr^{-1}}$) that stochastic sampling of the
  IMF is not expected to lead to the formation of any star with
  $M\gtrsim 30~\mathrm{M_{\sun}}$ within the typical dynamical time of
  a molecular cloud ($t_{dyn}\sim10^4$~yr). 

  Since the observed UV luminosity is significantly affected by dust
  extinction, we include in our modeling dust extinction with an
  empirically calibrated formula following \citet{bouwens2013} and
  \citet{meurer99}, which link extinction to the UV-continuum slope
  $\beta$ (for a spectrum modeled as
  $f_{\lambda}\sim\lambda^{\beta}$): $A_{\mathrm{UV}}=4.43+1.99\beta$.
  We model the observations by \citet{bouwens2013} as:
\begin{eqnarray}
& & <\beta(z,M_{AB})>= \\ 
& & \nonumber
\begin{cases}
a(z)\exp{(b(z) M_{AB})} + c & \text{if } M_{AB} \geq M_0 \\
c - a(z)\exp(b(z) M_0)(-1+b(z)(M_0-M_{AB})) & \text{if } M_{AB} < M_0 \\
\end{cases}
\end{eqnarray}
where $c=-2.276$, $M_0=-22$ and $a(z)$, $b(z)$ are determined from fitting
the \citet{bouwens2013} measurements. Assuming (like in
\citealt{tacchella13}) a Gaussian distribution for $\beta$ at each
$M_{\mathrm{UV}}$ value with dispersion $\sigma_{\beta}=0.34$, then
the average extinction $<A_{\mathrm{UV}}>$ is given by:
\begin{equation}
<A_{\mathrm{M_{UV}}}>=4.43+0.79\ln(10)\sigma_{\beta}^2+1.99<\beta>.
\end{equation}
This dust extinction framework differs slightly from the one we
adopted earlier where we had a linear dependence for
$\beta(M_{AB})$. In fact, the latest determination of $\beta$ by
\citet{bouwens2013} shows evidence for a curved relation between UV slope
and absolute magnitude. We adopt empirically an exponential function,
which provides a good fit, and then extrapolate it linearly, like
in our previous work, at high luminosities ($M_{AB}<-22$, where there
are only limited observations). The use of the exponential fit for
faint magnitudes has the advantage of avoiding unphysical negative
dust corrections.

To model the production of GRBs, we include a metallicity-dependent
efficiency as in \citet{trenti2013}. Galaxies are assigned a
metallicity dependence on redshift and luminosity following the
relation derived by \citet{maiolino08}, after we convert our UV
luminosity in stellar mass. Specifically, we use their Equation (2)
with coefficients given in their Table 5 for the \citet{bruzual03}
spectral energy distribution template. Coefficients are linearly
interpolated in redshift space among data points. At $z>3.5$ we assume
no further evolution of the mass-metallicity relation. {\rev Since our
  model implies a one-to-one map of stellar mass to star formation
  rate at fixed redshift, with the assumption of a mass-metallicity
  relation we are also defining a metallicity-star-formation rate
  relation. In Section~\ref{section:metal_outcome} we compare our model
  results to the fundamental relation derived by
  \citet{mannucci2010} at $z<2.5$. }

The efficiency of GRB production versus metallicity $Z$ is based on
the idea that we expect two main contributions to long-duration GRB
production. A channel broadly based on evolution of single massive
stars, where metallicity plays a crucial role in regulating mass loss
via winds (the Collapsar model, e.g. \citealt{yoon06}), plus
alternative channels without strong metallicity dependence (e.g., for
binary progenitors, \citealt{fryer05}). Combining the output from the
\citet{yoon06} Collapsar simulations with a metallicity-independent
plateau, we write the total GRB efficiency $\kappa(Z)$
as:
\begin{equation}\label{eq:metal_eff}
  \kappa(Z)=\kappa_0\times\frac{a\log_{10}{Z/Z_{\odot}}+b+p}{1+p},
\end{equation}
where $\kappa_0$, $p$ $a$, and $b$ take the same values as in
\citet{trenti2013}: For $Z/Z_{\odot}\leq10^{-3}$, $a=0$, $b=1$; for
$10^{-3}\leq Z/Z_{\odot}\leq10^{-1}$, $a=-3/8$, $b=-1/8$; for
$10^{-1}\leq Z/Z_{\odot}\leq1$, $a=-1/4$, $b=0$; for $Z/Z_{\odot}>1$,
$a=0$, $b=0$. A fiducial value of $p=0.2$ is used to construct our
reference model. $\kappa_0$ is an overall normalization which has no
impact on predictions for the luminosity functions of GRB hosts. The
quantity $p/(1+p)$ can be interpreted as the probability that a GRB
originates from the metal-independent channel rather than from a
collapsar, \emph{in the limit of metallicity of the host galaxy
  approaching zero}. We emphasize that $p$ is \emph{not} a relative
probability but rather a minimum, metal-independent \emph{plateau}
value for the efficiency of forming GRBs. The relative probability of
having GRBs originating from the two channels can only be computed
after taking into account the metallicity distribution of the star
forming galaxies, and this quantity generally strongly depends on the
redshift because of the evolving mass-metallicity relation of galaxies
(see Section~\ref{sec:results} for a more detailed discussion).

\begin{figure}
\begin{center} 
\includegraphics[scale=0.36]{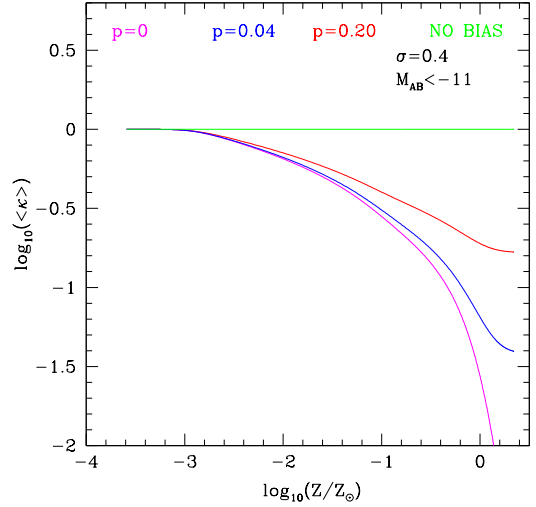}
\end{center}
\caption{Efficiency of GRB production per unit stellar mass as a
  function of host-galaxy metallicity for different values of the
  metal-independent plateau parameter $p$, from $p=0$ (magenta) to
  $p=+\infty$ (green).}\label{fig:kappa}
\end{figure}

To take into account both the intrinsic scatter in the
mass-metallicity relation, as well as the likely presence of a spread
in the metallicity of star forming gas within a given galaxy, we
assume that $Z$ follows a log-normal distribution, with $\langle ln(Z)
\rangle$ given by the mass-metallicity relation and $\sigma =
0.4$. This value corresponds to about 0.15 dex of intrinsic scatter in
$log10(Z)$ (e.g., see \citealt{Panter08}) and gives the resulting
$\langle \kappa(Z) \rangle$ shown in Figure~\ref{fig:kappa}.  We
calculate the comoving GRB rate from the model star formation rate,
weighted by $\langle \kappa(Z) \rangle$.

\begin{figure}
\begin{center} 
\includegraphics[scale=0.36]{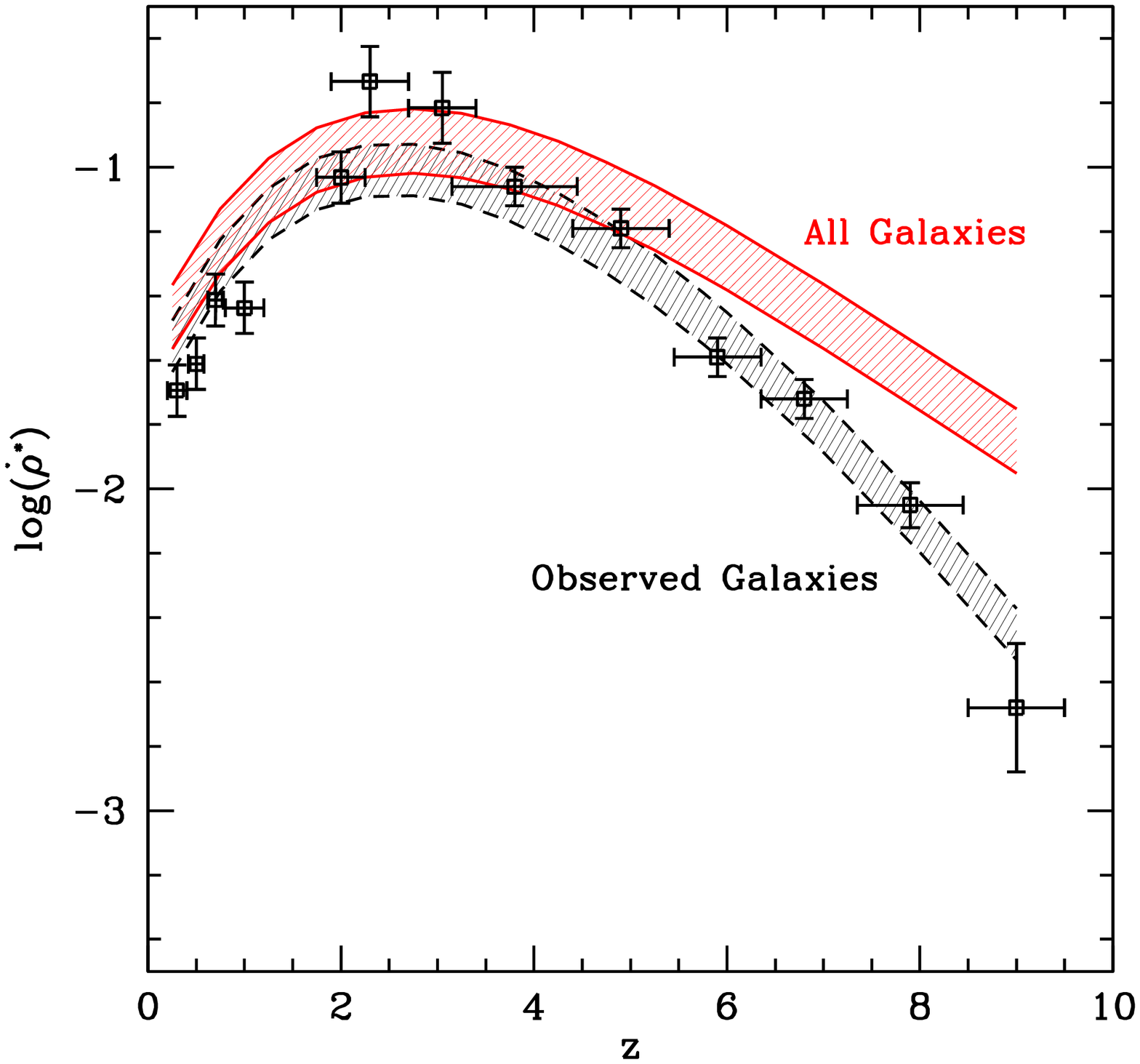}
\includegraphics[scale=0.36]{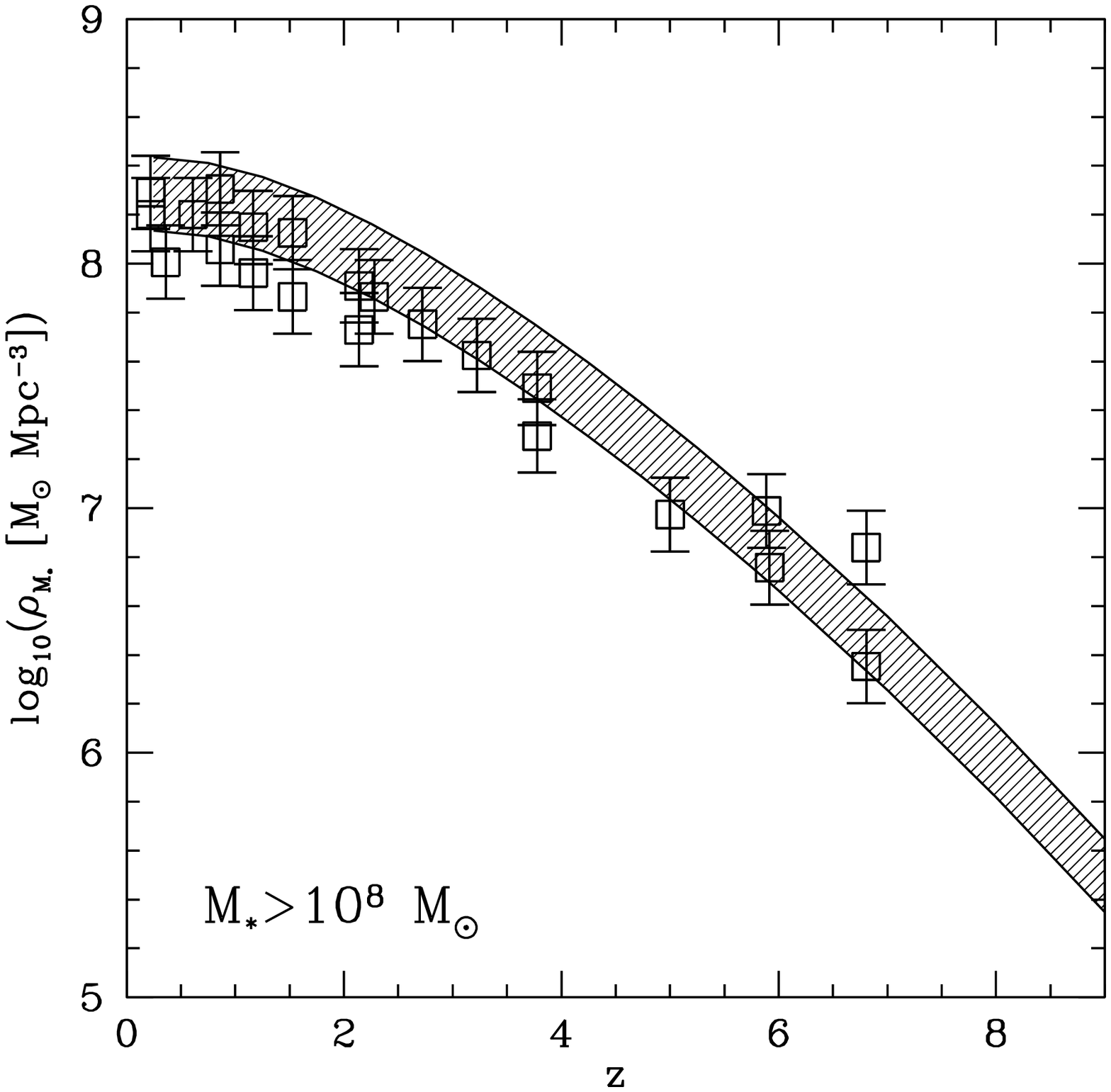}
\end{center}
\caption{Left panel: Star formation rate versus redshift (black
    points), inferred from LBG observations in the UV integrated to
    $M_{AB}=-17.0$ including dust corrections
    \citep{bouwens2013}. Predictions from our LF model are shown as
    black shaded area when integrated to the same limiting magnitude
    as the data, highlighting the ability of our framework to describe
    the star formation history of the Universe. The red shaded region
    shows model predictions when integrated to $M_{AB}=-11.0$
    (faintest galaxy assumed in the modeling).  Right panel: Stellar
    mass density versus redshift for a compilation of observations
    from \citet{tacchella13} (black datapoints) and predictions from
    our model. Both data and model predictions for stellar mass
    density are integrated for
    $M_*>10^8~\mathrm{M_{\sun}}$.}\label{fig:rates}
\end{figure}

To demonstrate that our framework is successful in describing the
evolution of the properties of the LBG population, we compare model
predictions for the star formation rate (luminosity density) and
stellar mass density to observations in Fig.~\ref{fig:rates}. Further
model validation and discussion of the comparison with observations of
the LBG luminosity functions and specific star formation rates from
$z\sim 0.3$ to $z\sim 8$ can be found in \citet{tacchella13}. 

To derive predictions for the UV LF of GRB host galaxies, we start
from the dust-attenuated (observed) LF of LBGs
$\phi_{LBG}(L_{obs})$, and apply a weighting to the LF which takes
into account the fact that the galaxies are selected based on the presence of a
GRB. First, since the GRB rate is proportional to the star formation
rate, more luminous host galaxies are preferentially present in a
sample of observations targeted at GRB locations. Second, the
metallicity bias must also be accounted for, introducing a further
weight by $\langle \kappa(Z(M_{AB}) \rangle$. When introducing these
weights, it is crucial to note that the GRB rate is proportional to
the intrinsic star formation rate and not to the observed
(dust-attenuated) one, therefore the correct weight to use is
$L_{int}=10^{-M_{AB}^{(int)}/2.5}$, where $M_{AB}^{(int)} =
M_{AB}^{(obs)}-A_{UV} (M_{AB}^{(obs)})$. Hence we can write:
\begin{equation}
\phi_{GRBhost} (L_{obs}) \propto \phi_{LBG}(L_{obs}) \times L_{int}  \times \langle
\kappa (Z) \rangle, 
\end{equation}
which can be rewritten as:
\begin{equation}\label{eq:phiGRBhost}
\phi_{GRBhost} (L_{obs}) \propto \phi_{LBG}(L_{obs})\times  L_{obs} \times
10^{A_{UV}(M_{AB}^{(obs)})/2.5} \times \langle \kappa (Z) \rangle. 
\end{equation}

Similarly, we proceed to construct predictions for the stellar mass
function of GRB hosts, {\rev and for their metallicity distribution.}

\section{Modeling result}\label{sec:results}

\subsection{Comoving GRB rate}

\begin{figure}
\begin{center} 
\includegraphics[scale=0.36]{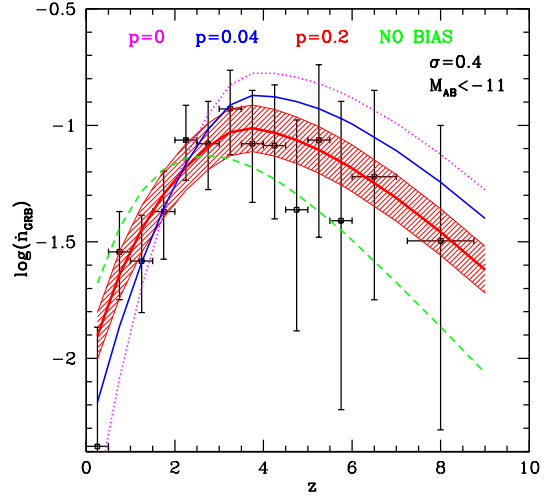}
\end{center}
\caption{ GRB comoving rate from \citet{wanderman10}, shown as black
  points with errorbars, compared to predictions of our models
  depending on the assumed efficiency of the metal-independent channel
  for GRBs. Our reference model ($p\sim0.2$, shown in bold red with
  shaded region for typical model uncertainty) provides the best
  description of the data. Models with stronger metal bias are shown
  in magenta and blue, while a no-bias model is shown in
  green.}\label{fig:rgb_rate}
\end{figure}

\begin{figure}
\begin{center} 
\includegraphics[scale=0.36]{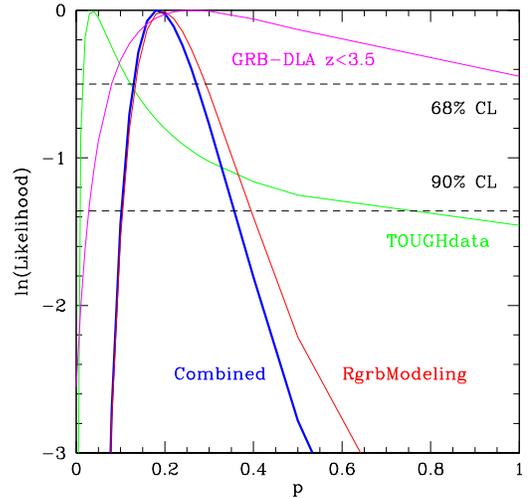}
\end{center}
\caption{Likelihood for the value of the metal-independent channel for
  GRB production ($p$), derived from the comoving GRB rate (red line),
  from the star formation rates of the TOUGH survey (unbiased
  sub-sample, green line), {\rev from the metallicity of GRB-DLAs at $z<3.5$
  (magenta line, observations from \citealt{cucchiara2014}), and
  combining the three constraints (blue line).} Dashed horizontal lines
  denotes the likelihood boundaries at 68\% and 90\%
  confidence. Because of the small sizes of the TOUGH unbiased
  subsample {\rev and of the GRB-DLA sample,} the strongest constraint on $p$
  originates from comoving rate modeling.}\label{fig:likelihood}
\end{figure}

The comoving GRB rate predicted by our model is shown in
Figure~\ref{fig:rgb_rate}, and compared to the observed event rate as
derived by \citet{wanderman10}. {\rev Since the rate encompasses
  information from the full sample of GRBs, it provides the most
  stringent constraints on the free parameter of our model, the
  metallicity plateau $p$. Thus we use the inference from the GRB rate
  to construct a canonical model, and then we test its predictions
  against observations of GRB hosts luminosity/star formation rate and
  metallicities, which are available only for small sub-samples.} As
expected, the data-model comparison is similar to our previous
analysis from \citet{trenti2013}. A small, but non-zero value of $p$
provides the best description of the redshift evolution of the GRB
rate. The highest likelihood when the rate is fitted at $z<6$ is given
by $p\sim 0.2$ (Figure~\ref{fig:likelihood}), which we assume as our
canonical model (red curve with shaded uncertainty region in
Figure~\ref{fig:rgb_rate}). Both lower and higher $p$ exhibit
systematic differences from the observations, implying that the
comoving rate points toward the presence of a metallicity bias in
GRBs, but that a non-zero fraction of events needs to originate from a
metal-independent channel. This conclusion confirms the findings of
\citet{trenti2013}, with a slightly revised quantitative result (the
canonical model considered in our earlier work had $p=0.3$) arising
because we updated the model calibration using the latest LBG
luminosity function and dust content determinations by
\citet{bouwens2013} and \citet{bouwens2014}.

\begin{figure}
\begin{center} 
\includegraphics[scale=0.36]{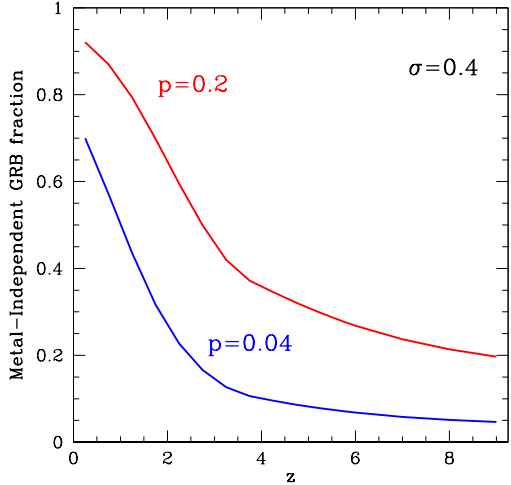}
\end{center}
\caption{Fraction of GRB produced by the metal independent channel as
  a function of redshift for our model with $p=0.2$ (solid
  red line), compared to the model with $p=0.04$ (solid blue line). At low
  redshift the majority of GRBs in both models are produced by this
  channel, while at high-$z$ the production approaches the asymptotic
  value $p/(1+p)$, and thus most GRBs are expected to originate from
  Collapsars.}\label{fig:channel}
\end{figure}

Figure~\ref{fig:channel} shows how the production of GRBs in our
canonical model switches from predominantly Collapsars (metal-biased)
at high-redshift, when most star forming sites have low metallicity,
to metal-independent (binary evolution) as the redshift
decreases. This happens because Collapsars are progressively
suppressed by the increasing metallicity while the metal-independent
channel continues to act unaffected, leading to an overall decrease
with redshift of the efficiency of GRB production per unit stellar
mass in star formation. By redshift $z=0$ our canonical model with
plateau $p=0.2$ predicts that over 90\% of the GRBs are produced by
the metal-independent channel (see Figure~\ref{fig:channel}). If we
assume instead a smaller value for $p$, for example $p=0.04$ in
Figure~\ref{fig:channel} (blue line), then we see that the relative
fraction of Collapsars is higher. In this case Collapsars are still
sub-dominant at $z=0$, but they quickly become the major mode of GRB
production at $z\gtrsim 1$, where the majority of observed GRBs are
located. {\rev Values of $p< 0.04$ further suppress the metal-independent
  channel, leading to a decrease of the rate of low-$z$ GRBs, while
  Collapsars become the dominant progenitor channel at all redshifts.}

\subsection{UV luminosity functions and star formation rates} 

\begin{figure}
\begin{center} 
\includegraphics[scale=0.36]{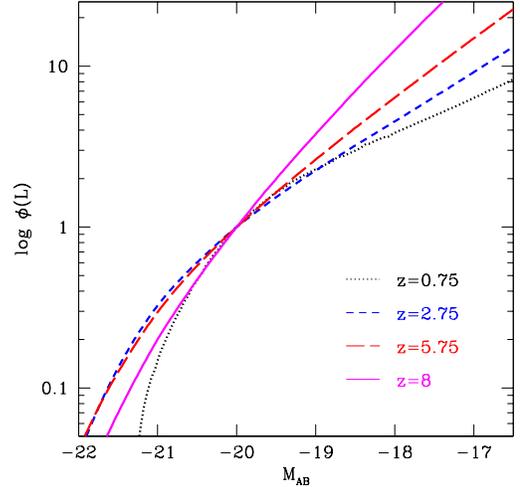}
\end{center}
\caption{Predictions for the UV luminosity function of GRB host
  galaxies based on our reference model which includes a GRB
  metallicity bias for different redshifts. The best fitting Schechter
  parameters associated to the curves shown are given in
  Table~\ref{tab:LFhost}. The curves have been normalized to have the
  same volume density at $M_{AB}=-20.0$. {\rev The plot shows that bright
  GRB host galaxies are rare both at low (black dotted, $z=0.75$) and
  very high redshift (solid magenta, $z=8$), respectively because of
  high metallicity (leading to both suppression of GRB production and
  significant reddening), and intrinsic rarity of massive, luminous
  galaxies at early times. Intermediate redshifts are shown as short
  dashed blue line ($z=2.75$) and long-dashed red line ($z=5.75$).} As
  the redshift increases, the LF becomes steadily steeper at the faint
  end.}\label{fig:LFhost}
\end{figure}

Using Equation~(\ref{eq:phiGRBhost}) we construct the GRB-host galaxy
luminosity function at different redshifts, and present the results in
Figure~\ref{fig:LFhost}. The best-fitting Schechter parameters
$M_{AB}^{(*)}$ and $\alpha$ are reported in Table~\ref{tab:LFhost} in
the magnitude range $-22.5\leq M_{AB} \leq -17.0$. {\rev The full
  luminosity function model output is published in the electronic
  edition of the article, and Table~\ref{tab:modeloutput} shows a
  portion for guidance regarding its form and content.} 

In general, we find that a Schechter LF provides a good description of
the modeling output in the magnitude range considered, with typical
values of $M_{AB}^{(*)}$ similar to those of the LBG LF at the same
redshift, that is $-21.2\lesssim M_{AB}^{(*)} \lesssim -19.4$, with
higher values at low $z$. The faint-end slope $\alpha$ is also
evolving with redshift, from $\alpha(z=0)\sim -0.1$ to
$\alpha(z=9)\sim -1.4$. Translating the luminosity functions into the
probability of detecting a GRB host galaxy as a function of the
limiting magnitude of the observations, we predict that GRB host
surveys reaching $M_{AB}=-18.0$ will have greater than 50\%
completeness at $z<4$ (see bold red dotted line in
Figure~\ref{fig:lf_distribution}). Observations reaching significantly
deeper, to $M_{AB}=-16.0$, are needed for the same completeness at
$z>7$, as a result of the steepening of the LF (see also
\citealt{salvaterra13} for independent modeling of the high-$z$
hosts).

\begin{figure}
\begin{center} 
\includegraphics[scale=0.36]{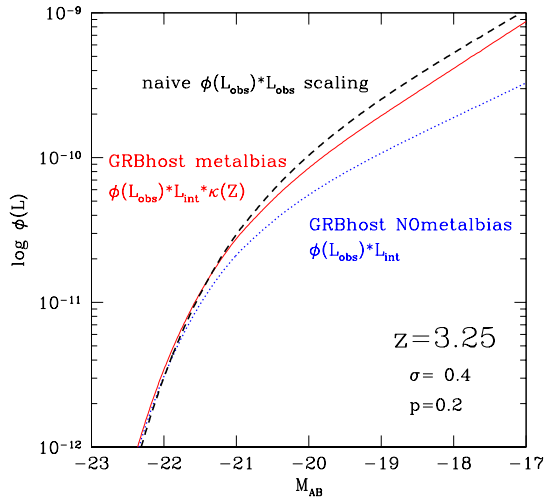}
\end{center}
\caption{Illustration of the combined effect of dust reddening and
  metallicity bias in determining the shape of the GRB host galaxy LF
  at $z=3.25$. The LF (red solid) is steepened by $\Delta \alpha\sim
  0.2$ because of metallicity bias in GRB production compared to a
  model where the GRB rate traces the star formation rate (blue
  dotted). Still, because of the difference between intrinsic versus
  observed luminosity, $L_{int}$ and $L_{obs}$, induced by reddening,
  the final GRB-host LF \emph{looks} very similar to
  $\phi(L_{obs})*L_{obs}$ shown as black long dashed line. Thus one
  could naively, but wrongly, infer that there is no metallicity bias!
}\label{fig:metal_dust_balance}
\end{figure}

When compared to the model predictions for the LF of LBGs, we see that
for most redshifts in our canonical case of $p=0.2$, the GRB hosts
empirically follow an approximate scaling with $\phi_{LBG}(L_{obs}) \times
L_{obs}$ \emph{when a metallicity bias is present}. This apparently
counter-intuitive result is illustrated in more detail in
Figure~\ref{fig:metal_dust_balance} and stems from the presence of
dust absorption. In fact, Equation~(\ref{eq:phiGRBhost}) shows that
since dust and metallicity correlate, the effect of the metallicity
bias is partially countered by the dust absorption term, present in
the equation because the GRB rate traces the intrinsic
(dust-corrected) luminosity density. This is a key finding of our
work: If one were to neglect the impact of dust and only consider the
observed LF of LBG, then there would be the expectation that observing
a GRB host luminosity function scaling as $\phi_{LBG}(L_{obs})\times
L_{obs}$ would mean that no metallicity bias is present.

Furthermore, because of the mass-metallicity relation of galaxies and
of its redshift evolution, the impact of the metallicity bias on the
LF shape depends strongly on the redshift. At very low redshift, a
significant fraction of GRBs {\rev in the canonical model} originates
from the metal-independent channel, since GRB production by collapsar
is suppressed in most hosts. In fact, the mass-metallicity relation
we use in this work, which is based on
\citet{kewley2008} at $z\sim 0$\footnote{Our mass-metallicity relation
  at $z=0.07$ derives from 
  \citet{kewley2008} with calibration done using the
  \citet{kewley2002} method, and conversion of stellar masses to
  account self-consistently for the \citet{bruzual03} initial mass
  function. We recall we are simply implementing the relation by
  \citet{maiolino08}, hence further details are available from that
  work (see, e.g. their Table 5 and Figure 7).}, predicts $Z\geq
Z_{\sun}$ (with $Z_{\sun}$ corresponding to $12+\log{(O/H)}=8.7$) for a
galaxy with stellar mass $\gtrsim 2\times 10^9~\mathrm{M_{\sun}}$
which has $M_{AB}\lesssim -17$ (after including dust
extinction). Then, at an intermediate redshift, there is a transition
to a regime where the effect of the metallicity bias is most
pronounced in producing a steepening of the GRB host LF. Finally, at
very high-$z$, the shape of the LF of GRBs is no longer affected by
the metallicity bias since the majority of star forming sites have
very low metallicities. {\rev At $z\gtrsim 5$ we thus predict that
  GRBs become essentially unbiased tracers of the star formation rate,
  since the majority of star forming sites have low metallicity and
  thus $\langle \kappa \rangle \to 1$ independent of the value of $p$
  (see Figure~\ref{fig:kappa}; see also \citealt{salvaterra13}).}

In Figure~\ref{fig:lf_distribution} we show predictions for typical
luminosities and star formation rates of GRB hosts for models with
different $p$.  If we set $p=0$ in Eq.~\ref{eq:metal_eff} (strong
metallicity bias), GRBs cannot be hosted in galaxies with metallicity
$Z\geq Z_{\odot}$, which introduces a sharp bright-end cut-off in the
LF. This is evident by looking at the sharp decrease in all luminosity
quantities {\rev at $z<3$ for the $p=0$ model (red lines in the top left
  panel of Figure~\ref{fig:lf_distribution})}.  At higher redshift, a
suppression of the host LF at the bright end is still present, but
hardly distinguishable from our canonical model. The $p=0$ case is
clearly an extreme scenario, already ruled out by observations of the
high metallicity of some GRB hosts (e.g., \citealt{levesque10c}), but
its analysis is still useful to highlight the impact of a strong
metallicity bias on the GRB host LF.

A scenario where the metallicity bias is absent and the GRB rate
traces the star-formation rate is difficult to discriminate from our
canonical model based on $z<1$ observations only, since the two
luminosity functions are essentially identical down to the median
luminosity of the GRB hosts. The strongest difference between the two
scenarios appears instead at $2.5\lesssim z \lesssim6$, when the median
host luminosities differ by about one magnitude.

This complex situation overall suggests caution when drawing
conclusions on metal-bias from a small sample of GRB hosts carrying
out a generic comparison with the luminosity and stellar mass
functions of star-forming galaxies. For example, \citet{jakobsson2005}
conclude that no metallicity bias is present, but they neglect the
dust impact discussed above. When cast in our framework, the GRB host
LF determined in that study would instead provide a weak-evidence in
favor of the presence of a metallicity bias at a strength broadly
consistent with our canonical model. In fact, we would expect that in
absence of metallicity bias the GRB-host LF is shallower by $\Delta
\alpha \sim 0.2$ at the intermediate redshifts analyzed by
\citet{jakobsson2005}.

\subsection{Stellar Masses of GRB hosts}

Figure~\ref{fig:stellar_masses} presents the predictions for the
stellar masses of GRB hosts based on our model. Qualitatively, the
same trends discussed above for the luminosity functions are present
and intermediate redshift hosts appear the most promising to
discriminate models with different $p$. This is not surprising, since
our model has by construction a one-to-one correspondence between
stellar mass and UV luminosity (even though the relation between mass
and luminosity changes both with redshift and halo mass because of the
evolution of the galaxy assembly time; see Section~\ref{sec:model} and
\citealt{tacchella13,trenti2013}). Thus, while our model describes well
the redshift evolution of the stellar mass density, as well as the
flattening at the low-mass end with decreasing redshift, as observed
for example by \citet{ilbert13}, our choice to start from UV
luminosity needs to be taken into account in a data model
comparison. In fact, this suggests to give more weight to analyses of
the star formation rates (UV luminosity), rather than the host mass
functions. We stress that this is true in general, and it is not
resulting just because we are using a simplified model: The GRB rate
traces recent star formation, and not the total stellar mass of the
host galaxy. No matter how massive and low-metallicity a galaxy is, if
its star formation rate is approaching zero, so will be its GRB
rate. {\rev Therefore, inference on the metallicity bias based on
  stellar mass functions, such as that of \citet{boissier2013}, might
  be affected by higher amount of noise compared to analyses based on
  star formation rate.}

{\rev Figure~\ref{fig:stellar_masses} shows that in case of a very
  strong metallicity bias ($p=0$), we expect a slight rise in the
  median stellar mass of the GRB hosts from $z=0$ to $z\sim
  2.5$. Models with weaker or no bias predict instead a steady
  decrease of the host stellar mass with redshift. When these model
  predictions are compared against observations (within the caveats
  discussed above), it is fundamental to take into account sample
  selection as well, since low-mass hosts may be preferentially missed
  at high redshift. In addition it is important to ensure modeling
  consistency as well (e.g. with respect to assumptions on the stellar
  initial mass function). A preliminary comparison with recent stellar
  mass measurements \citep{perley13,hunt2014} highlights that further
  modeling and data comparison is needed. In fact, the data, if taken
  at face value, show an increase of the median stellar mass of GRB
  hosts with redshift, which would indicate the presence of a strong
  metal bias in our model. However, the observed stellar masses are
  generally too high to be consistent with $p\sim 0$, and would rather
  be suggestive weaker metal bias (plus incompleteness at the low-mass
  end).}

\subsection{Metallicity of GRB hosts}\label{section:metal_outcome}

Finally, Figure~\ref{fig:metallicity_distribution} illustrates the
model predictions for the metallicity distribution of the GRB
hosts. Models with different $p$ have a qualitatively similar trend:
host metallicities are expected to decrease with increasing redshift,
simply reflecting the underlying mass-metallicity relation that we
assume in our model. The detailed choice of $p$ influences the shape
of the metallicity distribution, and how rapidly it evolves with
redshift. Interestingly, the upper tail (top 5\%) of the distribution
is very similar in all cases from $p=0$ to $p=+\infty$, implying that
using the maximum metallicity observed for a GRB host bears little
insight into the presence, or absence, of a metallicity bias. The
median, and the lower 20\% of the distribution have instead markedly
different behaviors, which should make it viable to differentiate between
models. For example there is a factor 10 difference in the value of
the bottom 20\% of the metallicity distribution going from $p=0$ to
$p=+\infty$ at $z=0$, with the relative difference in metallicity
remaining almost unchanged at high-$z$. 

{\rev The predictions for the metallicity of the GRB hosts have a
  direct dependence on the assumed relation between
  stellar-mass/star formation rate and galaxy metallicity. As
  discussed in Section~\ref{sec:model}, we use the redshift-dependent
  mass-metallicity relation of \citet{maiolino08}. An alternative
  possibility would have been to implement the fundamental relation
  observed at $z\lesssim 2.5$ by \citet{mannucci2010} which links all
  three quantities together without introducing any redshift
  dependence (at least in the redshift range considered by the
  authors). In Figure~\ref{fig:mannucci_vs_maiolino} we compare the
  metallicity as a function of star formation rate predicted in our
  model considering the two relations. It is very interesting, and
  reassuring, to note that at low redshift ($z=0.25$ shown in the
  figure), the two different assumptions lead essentially to the same
  relation. As the redshift increases, there is however a marked
  difference: the \citet{mannucci2010} fit remains at fairly high
  metallicity, while with the \citet{maiolino08} relation, we predict
  values that are lower by $\sim 0.5$ dex for high star formation
  rates (massive and luminous galaxies), with the difference growing
  significantly for lower star formation rates. Taken at face value,
  the \citet{mannucci2010} relation is inconsistent with the observations
  of low metallicities ($Z\sim 10^{-2}~Z_{\sun}$) in GRB-DLAs at
  $z\sim 2$ which are reported by \citet{cucchiara2014}, since it only
  predicts $Z\gtrsim 10^{-1}~Z_{\sun}$. In addition, it is not clear
  how to extrapolate the relation above $z>2.5$, since galaxies at
  higher redshift start showing systematic deviations, as
  \citet{mannucci2010} highlight in their Figure 4 (right
  panel). While these issues prevent us from implementing an analysis
  of GRB host observations with the \citet{mannucci2010} relation, our
  Figure~\ref{fig:mannucci_vs_maiolino} may suggest that the
  extrapolation of the mass-metallicity relation by
  \citet{maiolino08} is too steep for faint galaxies, possibly
  introducing systematic uncertainty in our analysis, which could be
  evaluated if large complete samples of GRBs metallicity
  measurements/limits were available.}

\section{A first application to GRB host observations}\label{sec:tough}

{\rev \subsection{Star Formation Rates}}

For a proper comparison between model and data, it is fundamental to
resort to complete and unbiased samples of follow-up
observations. Otherwise, systematics that are hard or even impossible to
quantify might affect the inferences obtained. For example, it is
likely to expect that observations of GRB hosts with short
observations leading to non-detections might be preferentially non
reported in the literature compared to a detection of a bright
host. This would lead to a luminosity function that is skewed toward
higher number density at the bright end, masking the presence of a
metallicity bias. Furthermore, dusty hosts might be preferentially
missed, and a high dust content might even compromise the success in
obtaining a redshift for the GRB from the afterglow, leading in this
case to an over-estimation of the impact of the metallicity bias. A
complete sample free of selection effects is therefore fundamental. 

This is the reason why we consider data from the optically unbiased
GRB host (TOUGH) survey \citep{hjorth12} to illustrate the potential
of a comprehensive data-model comparison as a tool to infer the
presence and strength of a metallicity bias in GRB production, {\rev
  and to test the inference derived from modeling of the GRB comoving
  rate}. Specifically, we restrict to the \emph{complete} sub-sample
presented in \citet{michal2012}, which includes UV-inferred star
formation rate {\rev (uncorrected for dust)} for all hosts and which
we overplot to our predictions in Figure~\ref{fig:lf_distribution},
{\rev which are also referring to \emph{observed} UV luminosity
  therefore providing an uncorrected star formation rate}. Despite the
low number of points (11 hosts) and their relatively low redshift
($z\lesssim 1.1$), the analysis of the observations provides a first
constraint on $p$ independent of the global GRB rate modeling. We
carried out a Maximum Likelihood analysis of the star formation rates
from \citet{michal2012} at varying $p$, and show the results in
Figure~\ref{fig:likelihood} (green line). The likelihood strongly
excludes $p=0$ and has a peak at $p=0.04$. Unsurprisingly, based on
the considerations of Section~\ref{sec:results}, there is a
near-plateau of high likelihood values at $p\gtrsim 0.04$: At 68\%
confidence, $0.01<p<0.12$, but the interval is broader and highly
asymmetric for the 90\% confidence region, with allowed values
$0.1<p<0.8$. This is the consequence of the limited discriminating
power of a low-$z$ sample. Yet, despite the large uncertainty, it is
re-assuring to see that $p$ values inferred independently from the UV
luminosity modeling and from the rate modeling (red solid) agree at
the $\sim 1.5\sigma$ level. When the two likelihoods are combined
together (blue solid), $p\sim 0.2$ remains the most likely
solution. {\rev This is essentially because a small sample of 11
  datapoints is not competitive against the larger amount of
  information encoded in the comoving GRB rate, derived from all the
  known events. A plateau value $p\sim 0.2$} implies a probability
$p/(1+p)\sim 0.15$ for GRBs to be produced by metal-independent
progenitors \emph{in the limit of very low metallicity}, while the
actual fraction of GRBs originating from the two channels as a
function of redshift is shown in Figure~\ref{fig:channel}.

{\rev \subsection{Metallicity Distribution}}

Observations of hosts metallicities can provide a {\rev further} test
to validate/reject the modeling outcome {\rev (which we show in
  Figure~\ref{fig:metallicity_distribution}) for different values of
  the efficiency parameter $p$.} Unfortunately, and unlike the case
for UV luminosity, data are not available for a \emph{complete}
sample, complicating a formal likelihood analysis.

Still, a qualitative comparison with results from the compilation of
GRB hosts with measured metallicities carried out by
\citet{savaglio2013}, {\rev and the new recent measurement/re-analysis
  by \citet{cucchiara2014}} yields good insight: Our canonical model
($p=0.2$, red lines in figure~\ref{fig:metallicity_distribution})
predicts {\rev hosts metallicities} that are too high compared to the
observed values {\rev at low redshift (green points in the figure).}

{\rev Values $p\lesssim 0.04$ would be needed to improve consistency
  with these data, pointing toward a lower efficiency of the
  metal-independent production channel, and thus a stronger presence
  of the metallicity bias, which has been argued by several studies in
  this redshift range \citep{savaglio09,castroceron10,perley13}, and
  also derived by our maximum likelihood analysis of the TOUGH host
  galaxy luminosity (green curve in
  Figure~\ref{fig:likelihood}). However, this in turn creates tension
  with the global modeling of the GRB rate, which prefers $p\gtrsim
  0.1$ at 90\% confidence.}

Of course, this {\rev inference on $p$ from host metallicity} is valid
only under the assumption that {\rev the observations we consider
  have} a distribution that is representative of that of a complete
sample. {\rev This is} complicated because of the difficulty of
measuring metallicities for GRB hosts both at the high end
(e.g. because of possible dust biases), and at the low end (since
hosts are intrinsically faint). {\rev The impact of observational
  biases in the likelihood analysis is clearly seen by analyzing
  subsamples of the measurements and upper limits on metallicity
  reported by \citet{cucchiara2014}, and shown as blue points in
  Figure~\ref{fig:metallicity_distribution}. If we restrict the
  likelihood analysis at $z\lesssim 3.5$ we have a good agreement
  between data and our canonical model, with $p=0.2$ providing a
  solution within the 68\% confidence interval (see magenta line in
  Figure~\ref{fig:likelihood}). However, inclusion of all data shifts
  the likelihood toward significantly higher $p$ values and absence of
  metallicity bias would be the preferred solution (see red line in
  Figure~\ref{fig:likelihood_appendix}). This possibly originates
  because $z>3.5$ spectroscopic observations have been carried out
  primarily at low-resolution, which provide lower limits to
  metallicity \citep{cucchiara2014}. On the other hand, if one
  restricts the sample only to high-resolution spectroscopy, then the
  likelihood analysis yields the opposite result, namely a
  preference for $p=0$ (blue line in
  Figure~\ref{fig:likelihood_appendix}). Still, even assuming this
  likelihood, when combined with the result from GRB rate modeling,
  the total analysis shifts only marginally away from our canonical
  $p=0.2$ model which remains within the combined 68\% confidence
  interval (black line in Figure~\ref{fig:likelihood_appendix}). }

For future progress, it would be extremely helpful to build an
observational sub-sample {\rev with uniform coverage} which is
complete and similar to the TOUGH dataset of star-formation rates
considered by \citet{michal2012}, but targeted at higher redshifts, in
order to obtain likelihood constraints comparable or even stronger
than those inferred from rate modeling. For this, host follow-up of
all GRBs presented by \citet{salvaterra12} would be ideal and could
solve the tension between the metallicity predictions which favor low
$p$, and the comoving GRB rate modeling which seems to favor higher
$p$.

\section{Conclusions}\label{sec:con}

\begin{figure}
\begin{center} 
\includegraphics[scale=0.34]{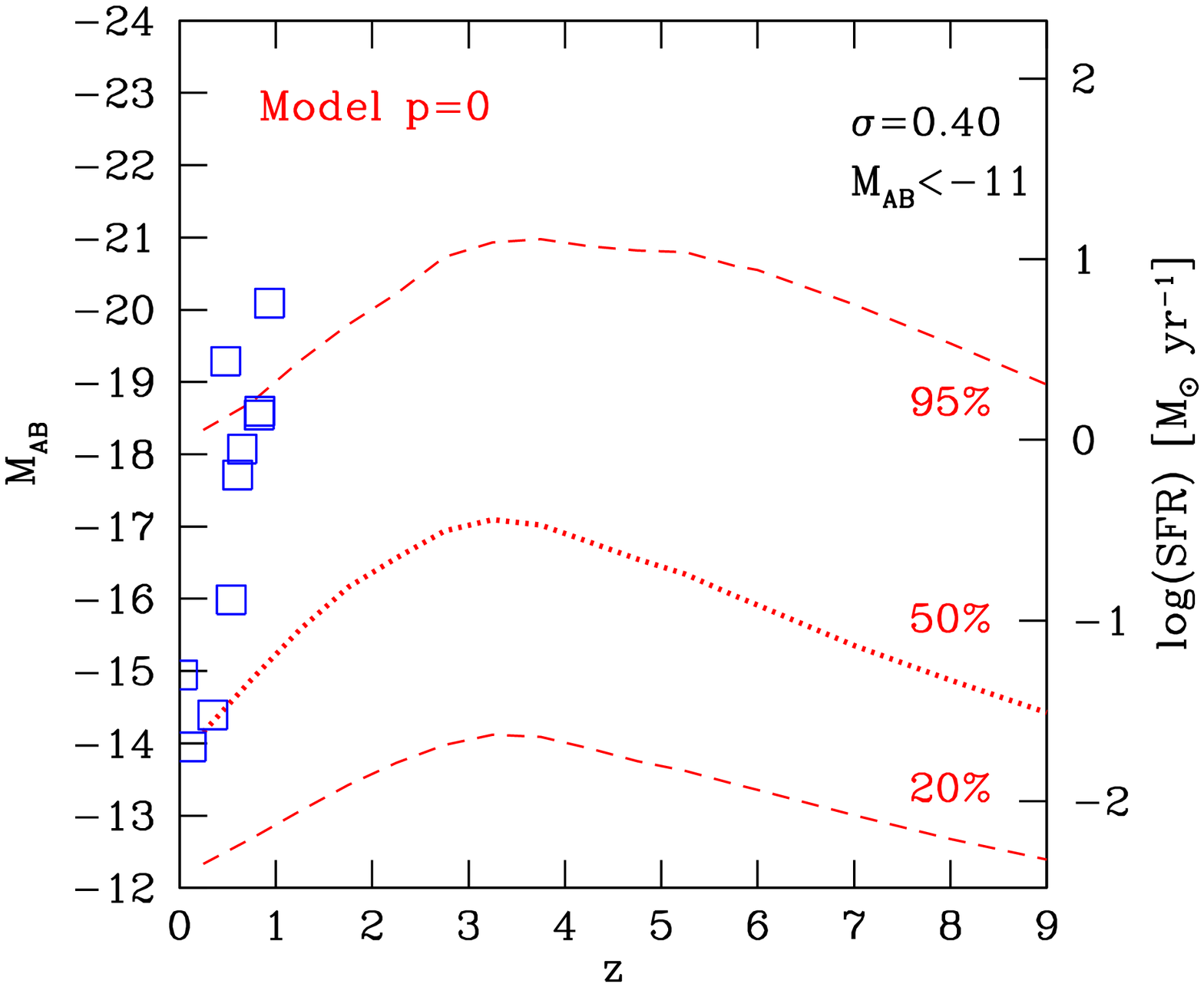}
\includegraphics[scale=0.34]{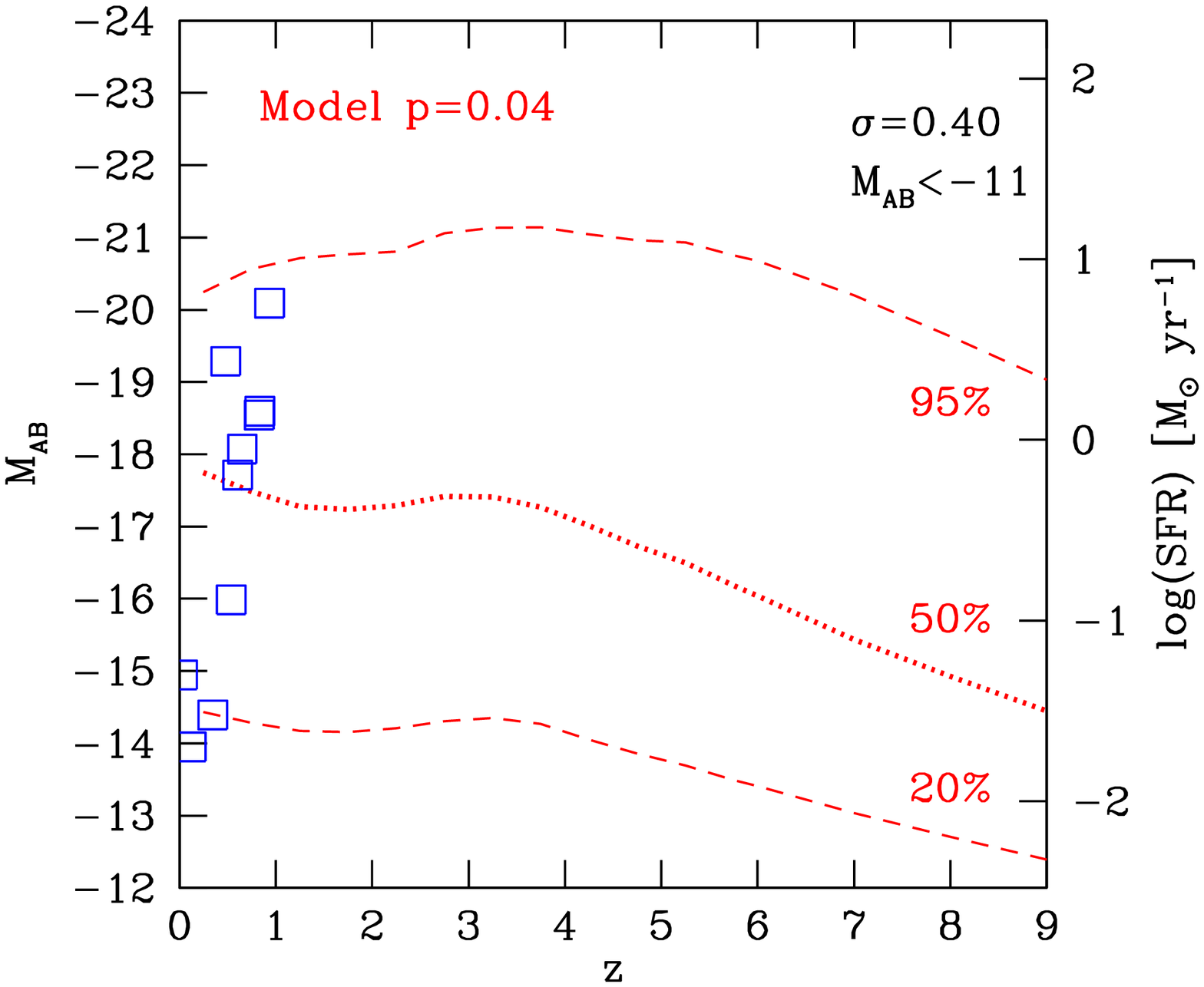}
\includegraphics[scale=0.34]{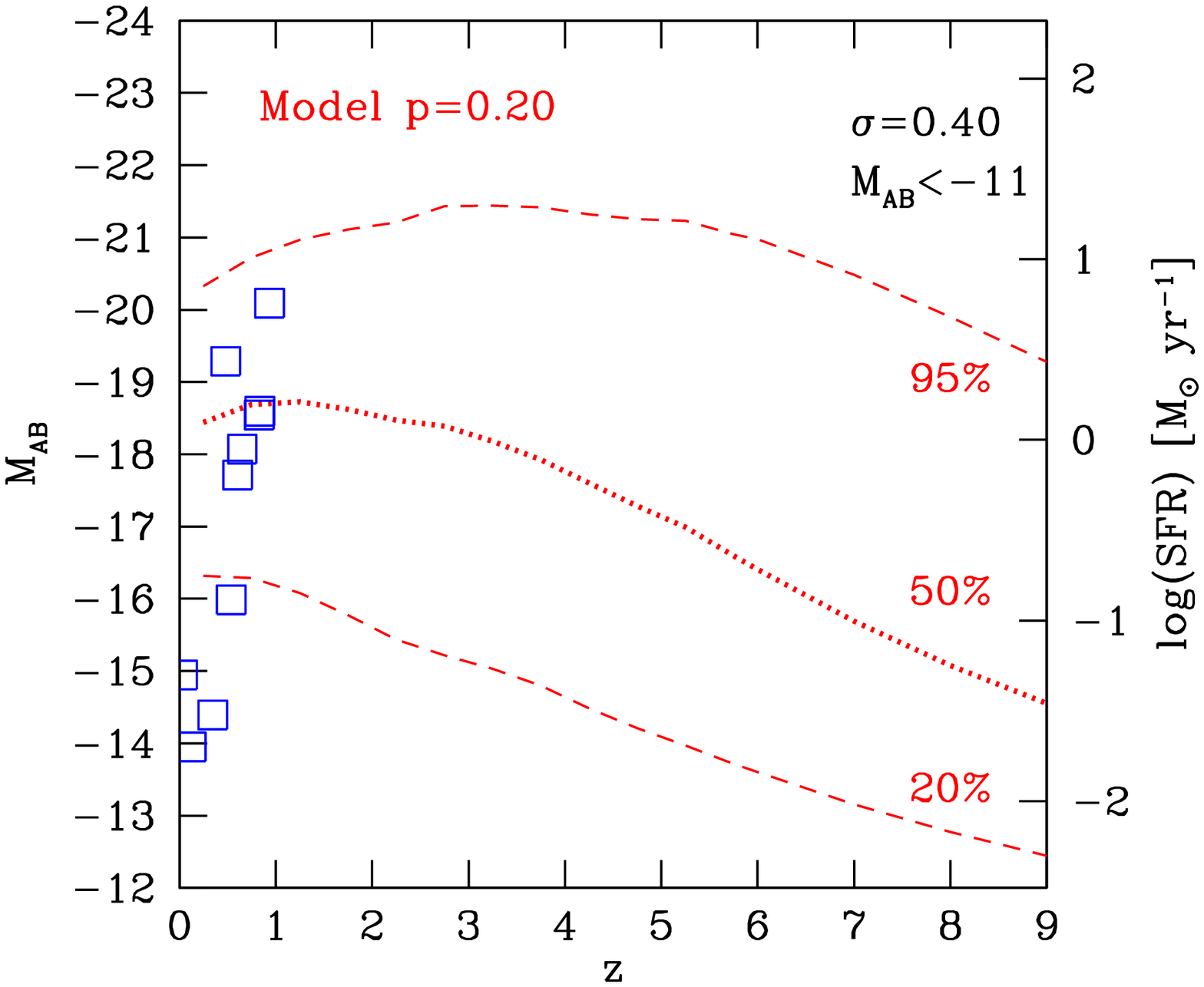}
\includegraphics[scale=0.34]{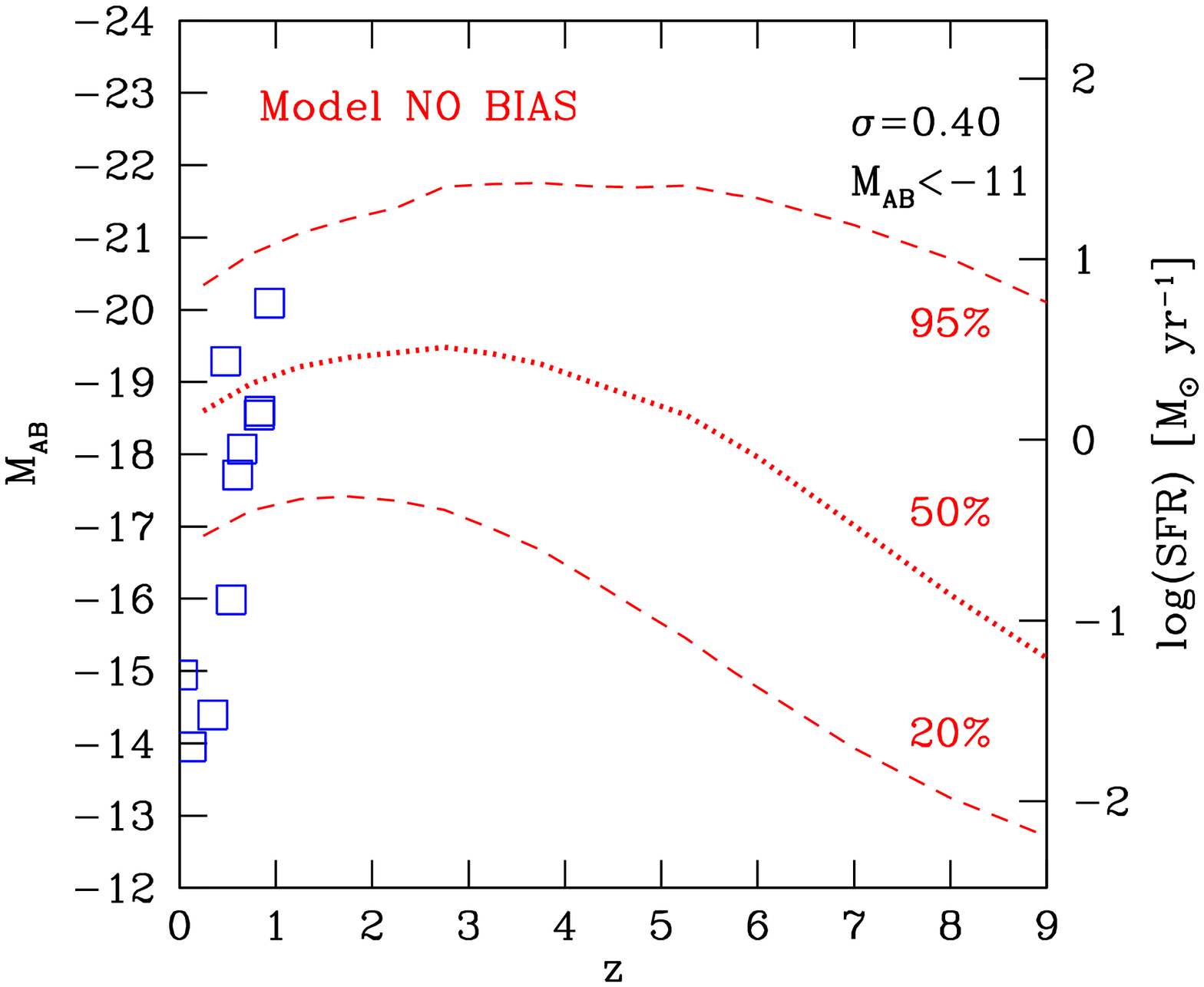}
\end{center}
\caption{{\rev Evolution of the upper 95\% (top dashed), median (bold
  dotted) and lower 20\% (bottom dashed) of the luminosity function
  (including dust reddening) of GRB host galaxies, with each panel
  showing model predictions for different efficiencies of the
  metal-independent channel for GRB production. Increasing values of
  $p$ are presented from top-left to bottom-right, from strong
  metal-dependence ($p=0$, top left) to absence of metal bias
  ($p=+\infty$, bottom right). Our canonical model $p=0.2$ is shown in
  the bottom left. For each panel, the AB magnitude (restframe UV at
  $1600$ \AA) is translated into a star formation rate on the right
  vertical axis, following \citet{madau98}. Data from the
  \emph{complete} TOUGH sub-sample of \citet{michal2012} are shown as
  blue open squares.}}\label{fig:lf_distribution}
\end{figure}

\begin{figure}
\begin{center} 
\includegraphics[scale=0.34]{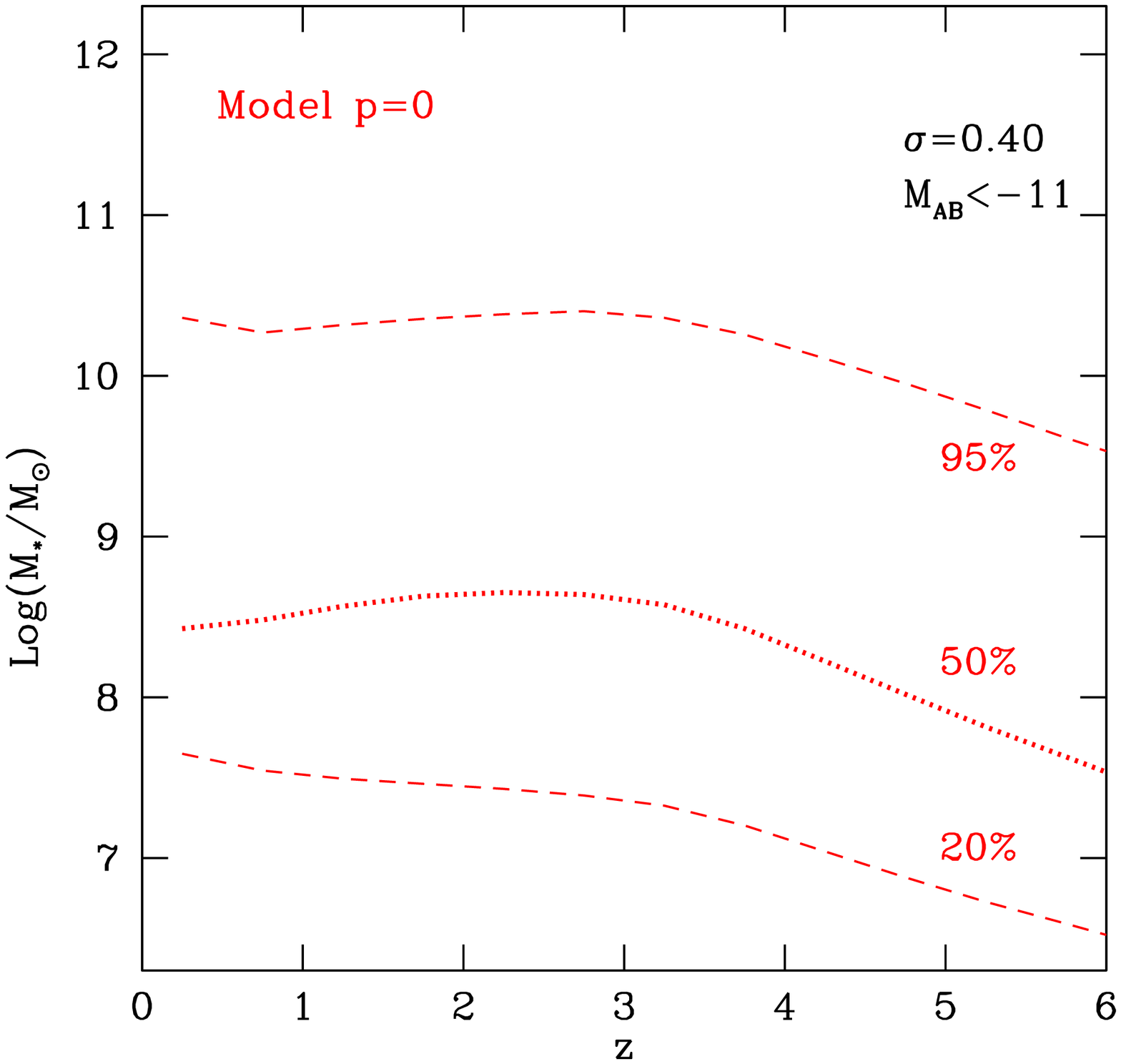}
\includegraphics[scale=0.34]{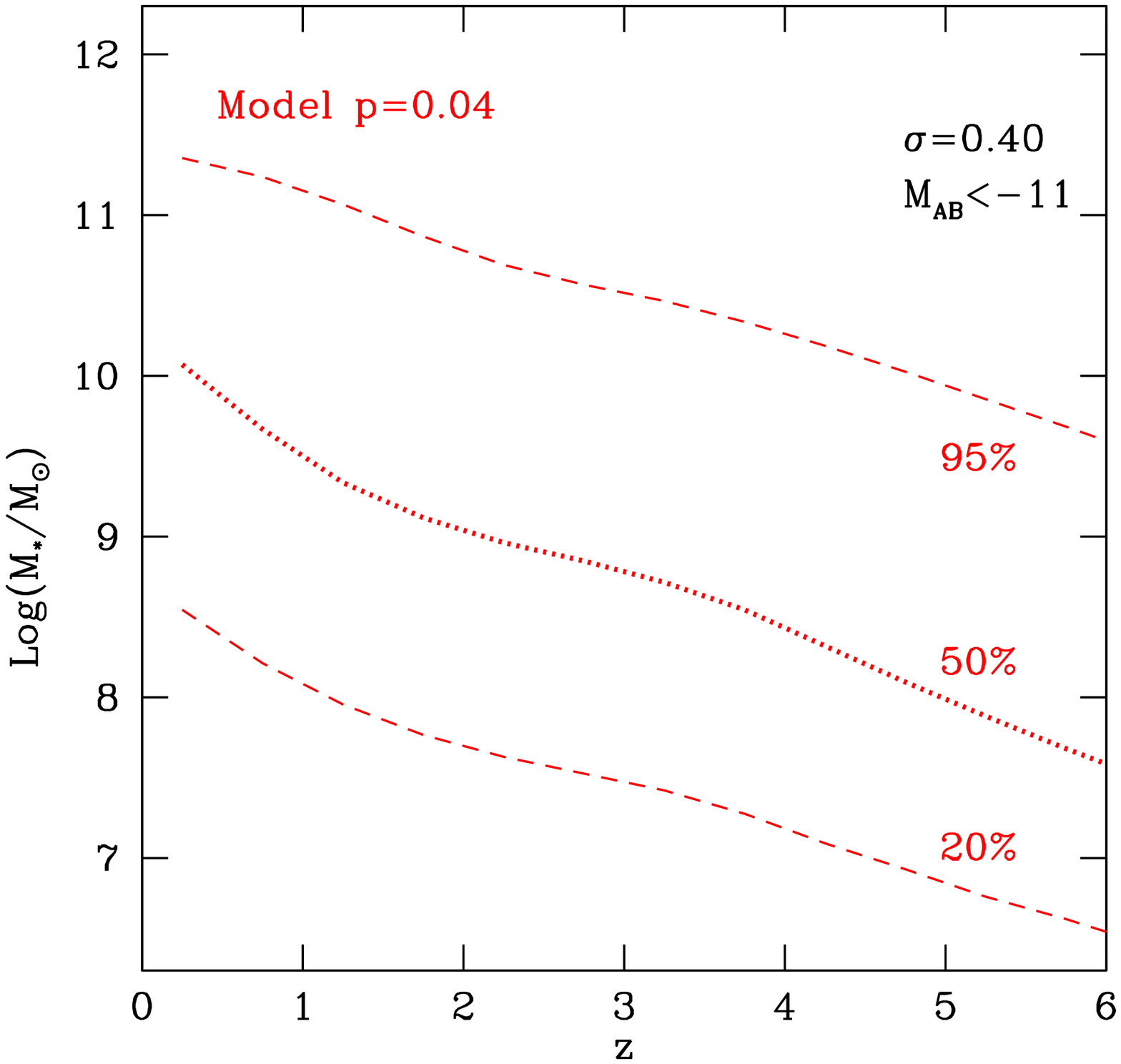}
\includegraphics[scale=0.34]{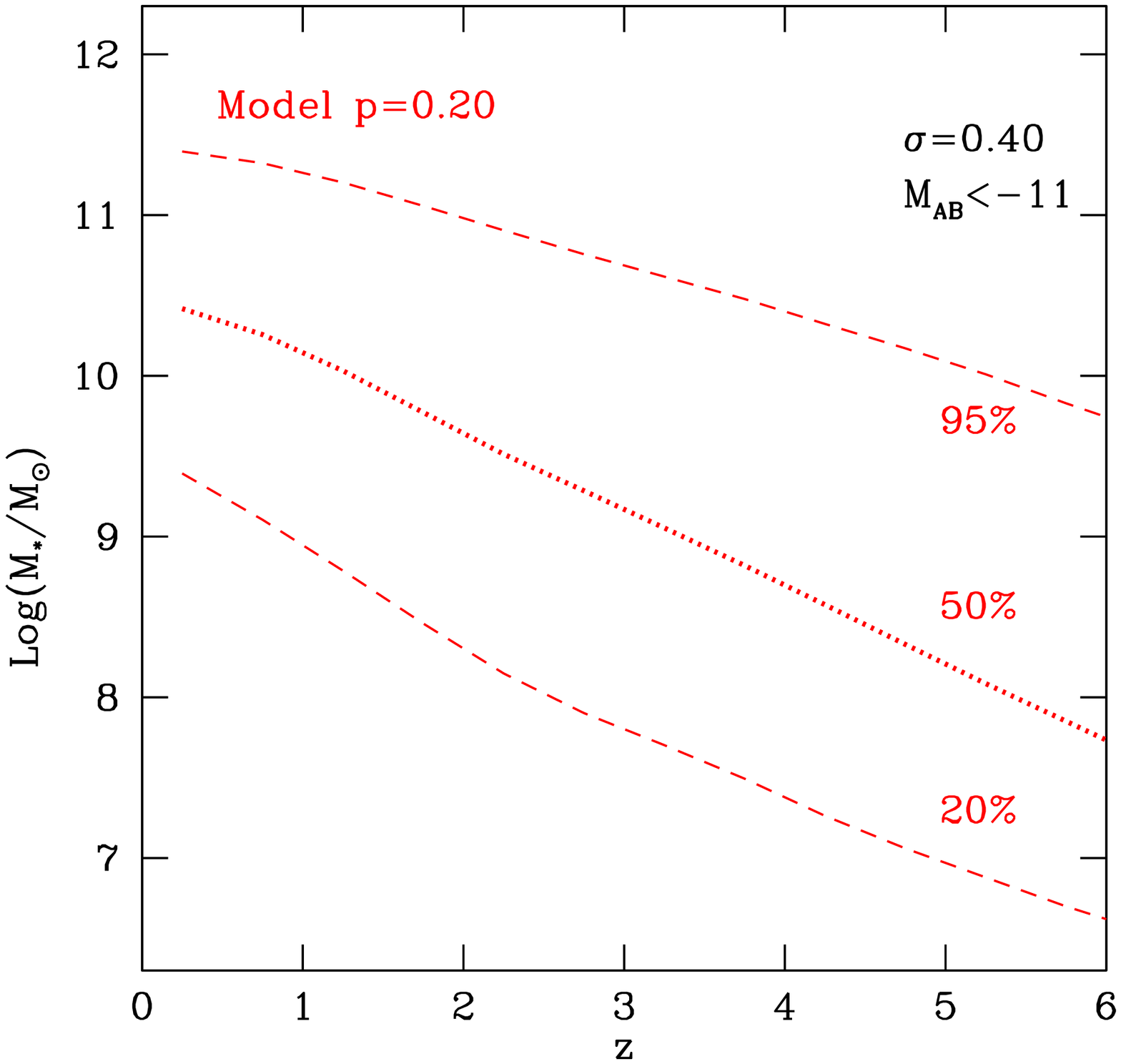}
\includegraphics[scale=0.34]{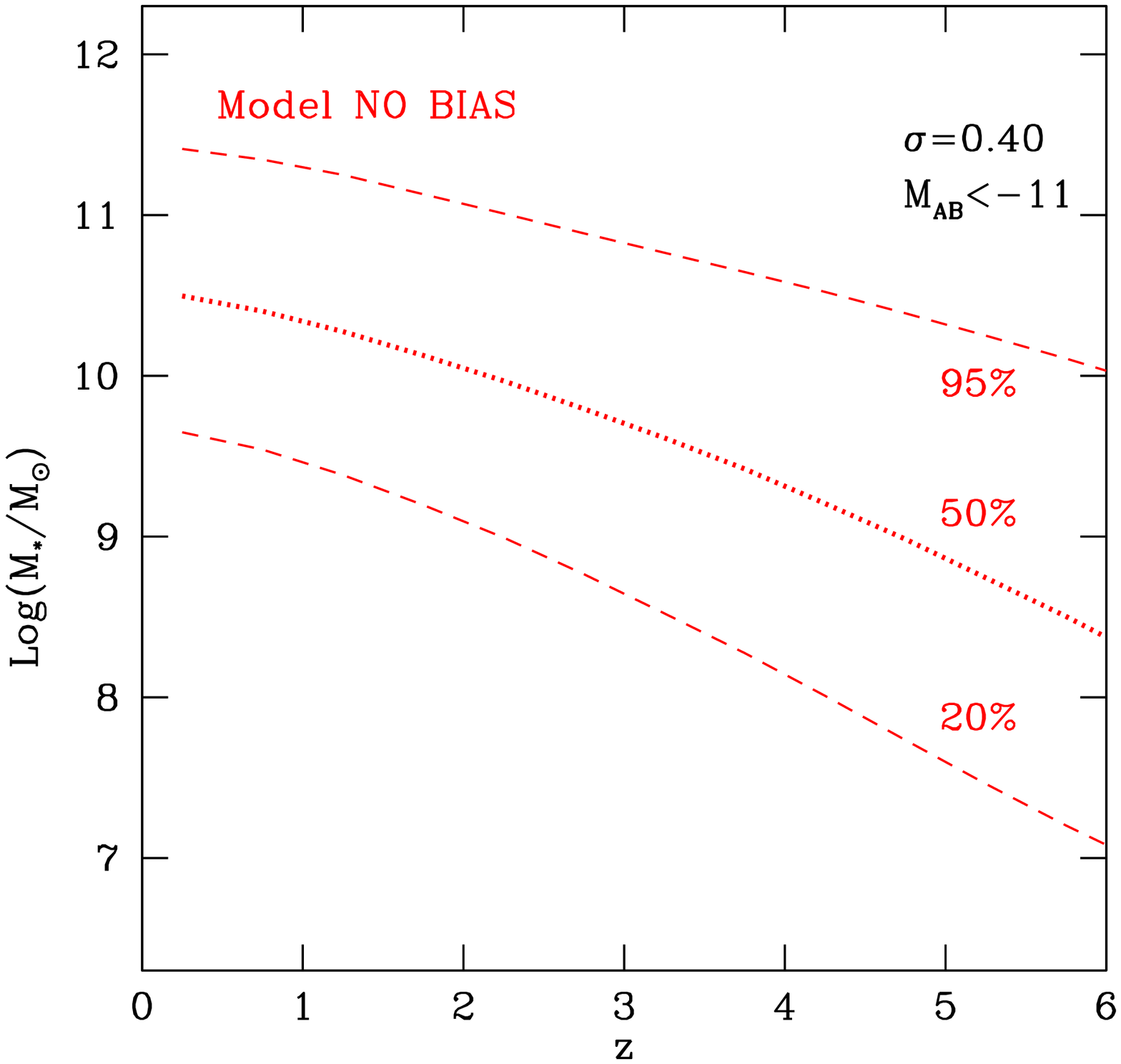}
\end{center}
\caption{{\rev Evolution of the upper 95\% (top dashed), median (bold
  dotted) and lower 20\% (bottom dashed) of the stellar masses
  of GRB host galaxies versus redshift. As in Figure~\ref{fig:lf_distribution}, each panel
  shows model predictions for different efficiencies of the
  metal-independent channel for GRB production.}}\label{fig:stellar_masses}
\end{figure}

\begin{figure}
\begin{center} 
\includegraphics[scale=0.34]{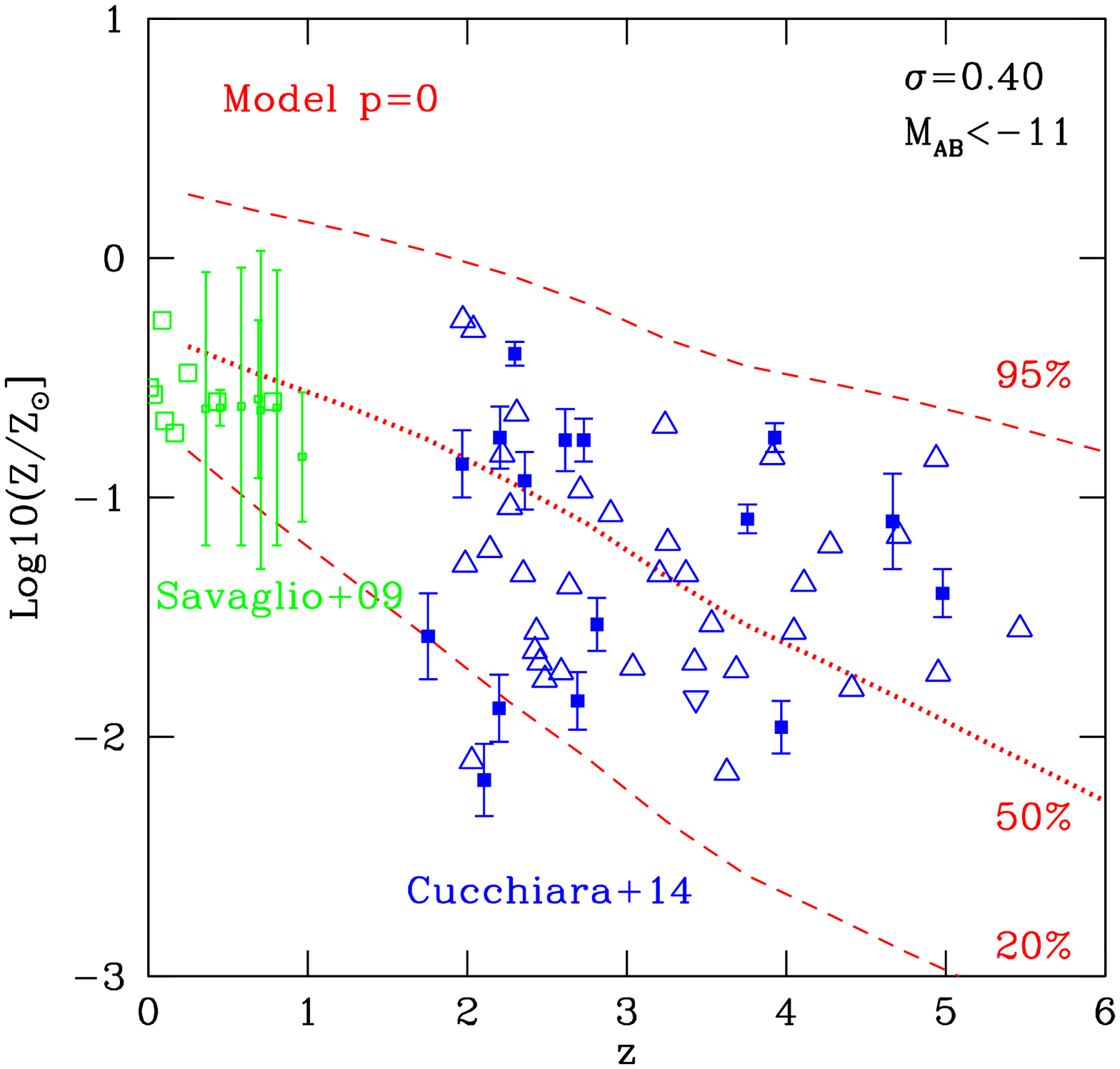}
\includegraphics[scale=0.34]{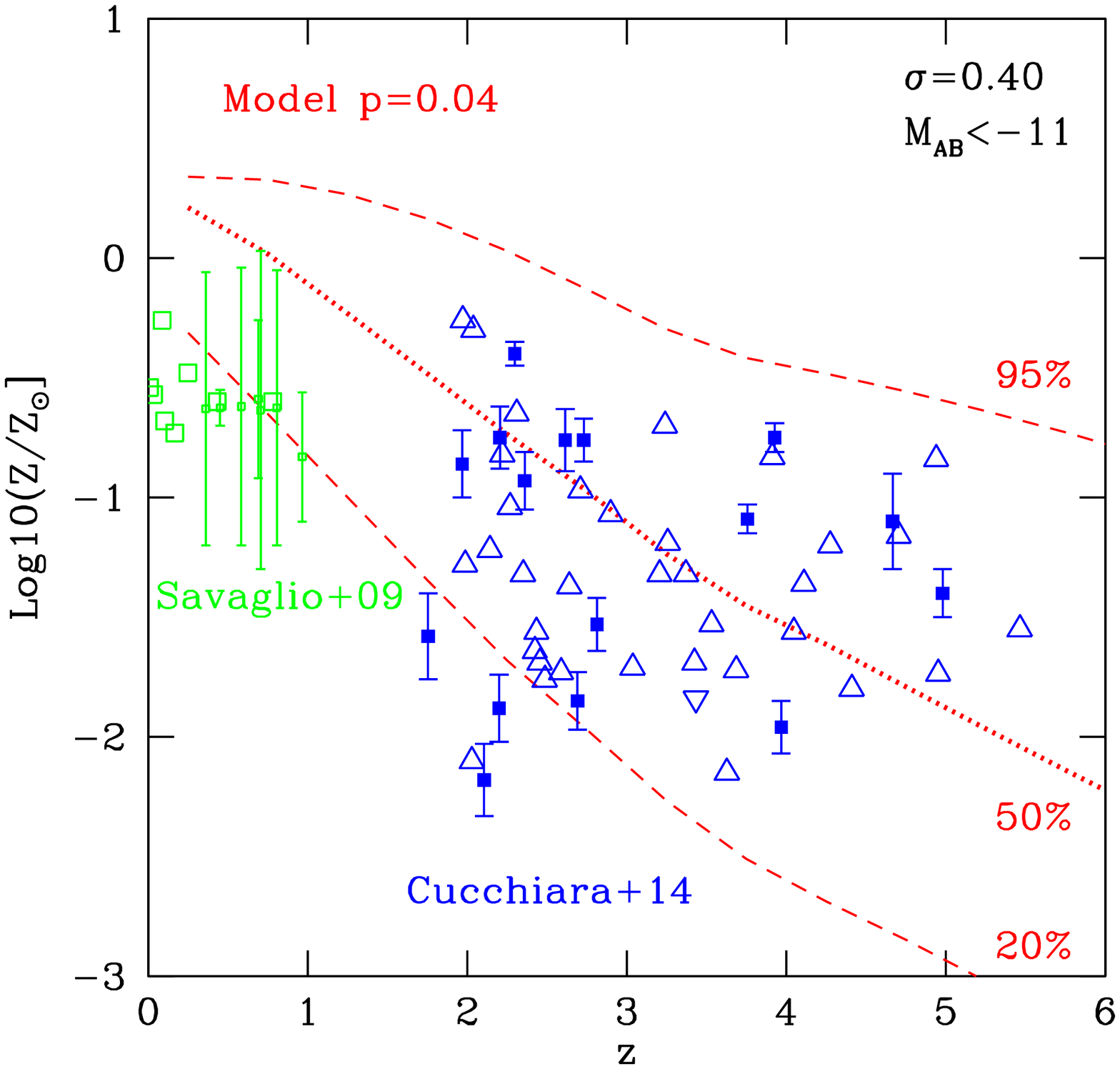}
\includegraphics[scale=0.34]{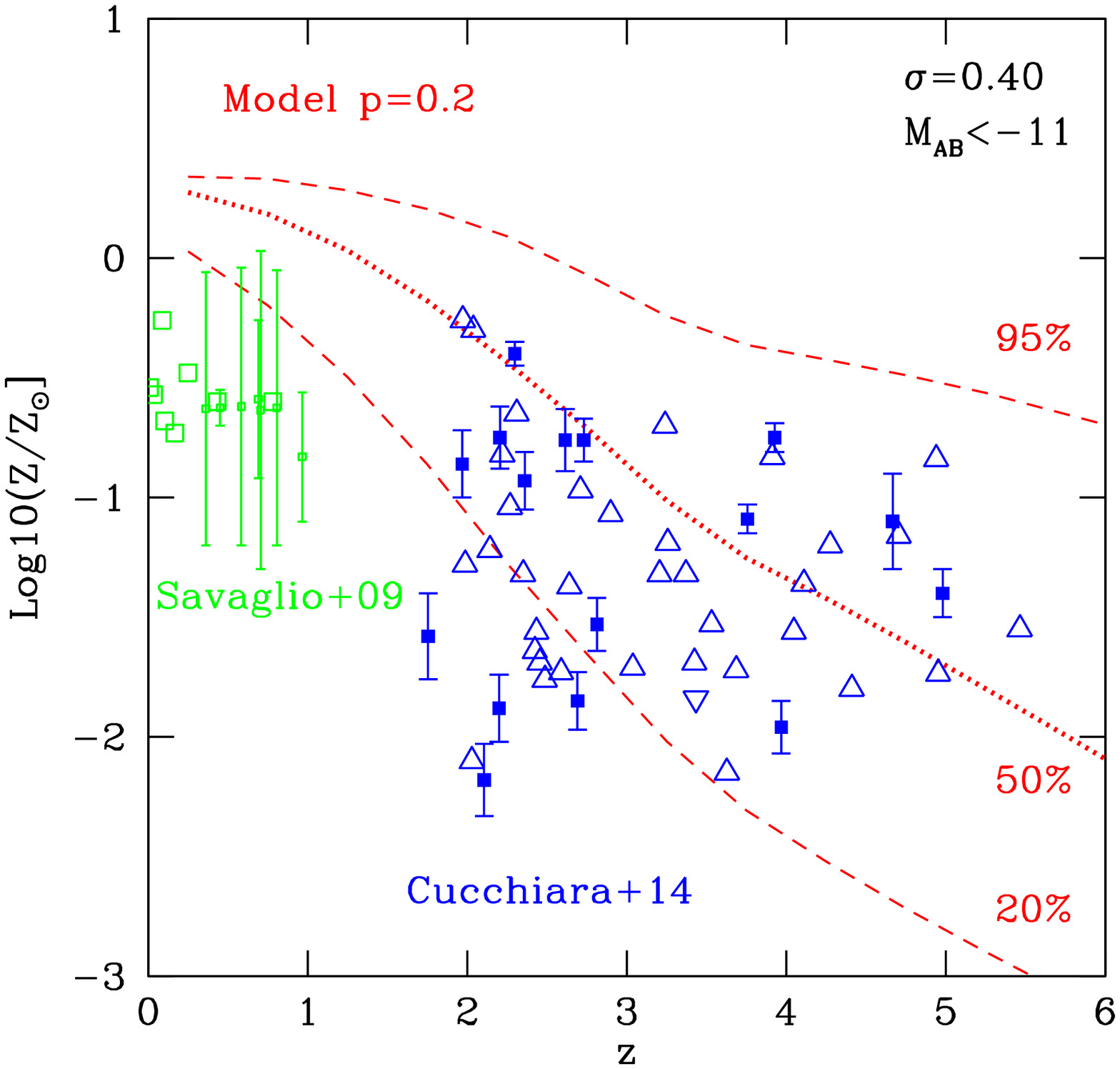}
\includegraphics[scale=0.34]{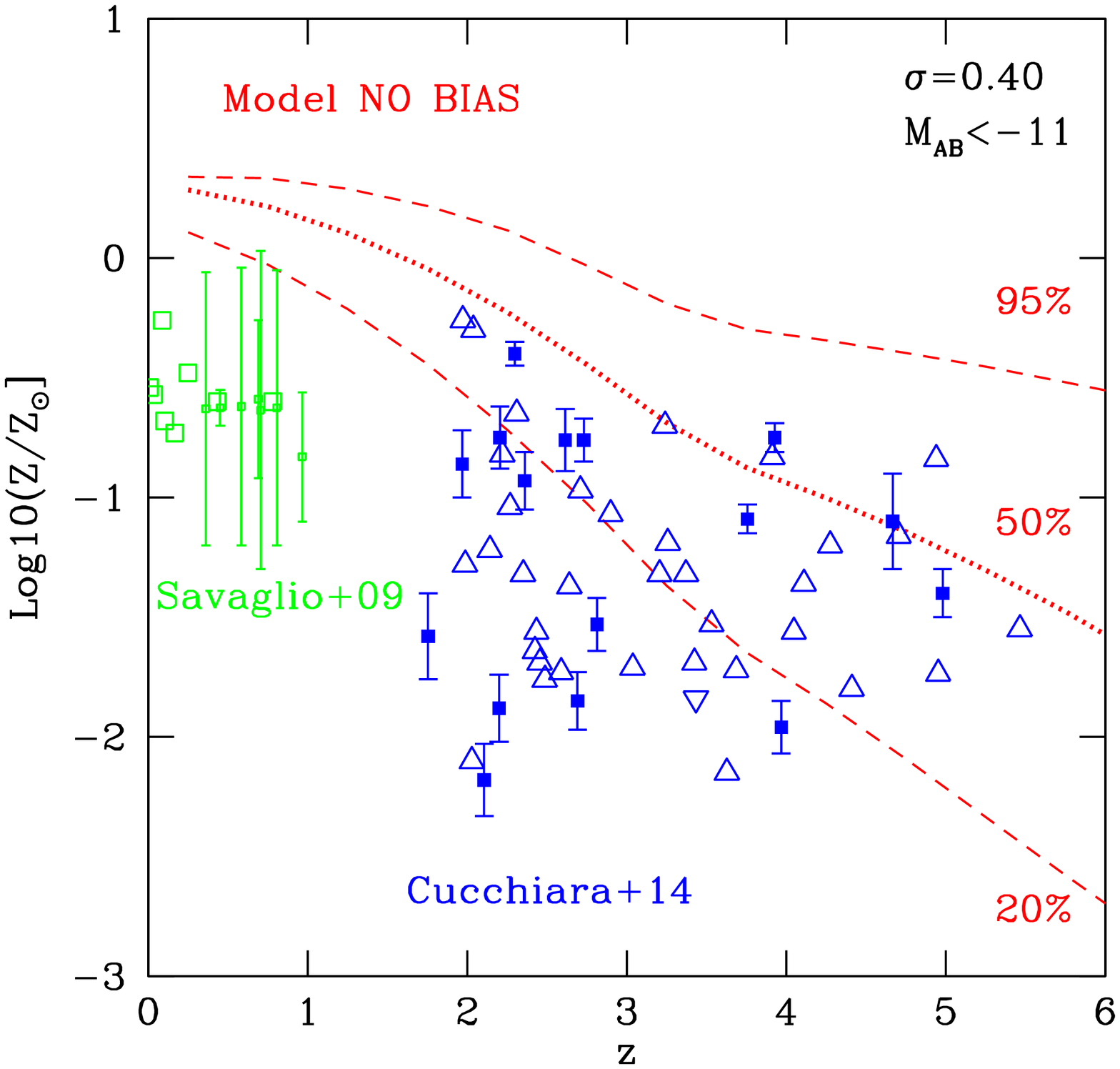}
\end{center}
\caption{{\rev Evolution of the upper 95\% (top dashed), median (bold
  dotted) and lower 20\% (bottom dashed) of the metallicity (solar
  units) of GRB host galaxies versus redshift. As in
  Figures~\ref{fig:lf_distribution}-\ref{fig:stellar_masses}, each
  panel shows model predictions for different efficiencies of the
  metal-independent channel for GRB production. Overplotted we
  report measurements and lower/upper limits on metallicity from
  \citet{savaglio09} and \citet{cucchiara2014}, although we stress
  that the samples are not complete, complicating the interpretation
  of the data-model comparison.}}\label{fig:metallicity_distribution}
\end{figure}

\begin{figure}
\begin{center} 
\includegraphics[scale=0.36]{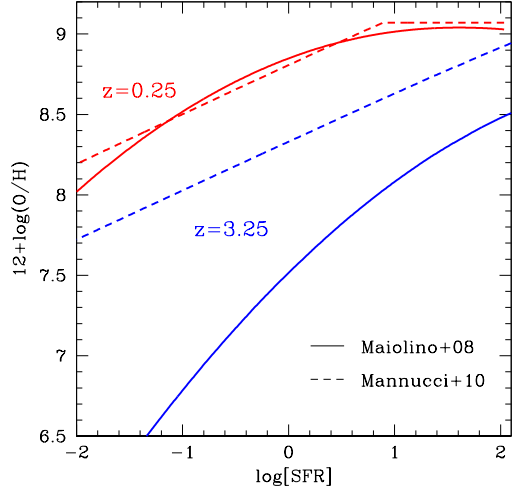}
\end{center}
\caption{{\rev Metallicity (in $12+log (O/H)$ scale with solar value associated
  to $8.7$) versus star formation rate as inferred through
  our model using the \citet{maiolino08} mass-metallicity relation
  (solid line) and the \citet{mannucci2010}
  mass-metallicity-star-formation-rate relation (dashed line) for two
  redshifts ($z=0.25$ in red) and ($z=3.25$ in blue). The two
  relations give comparable results at low $z$ but differ
  systematically at high $z$.}}\label{fig:mannucci_vs_maiolino}
\end{figure}

\begin{figure}
\begin{center} 
\includegraphics[scale=0.36]{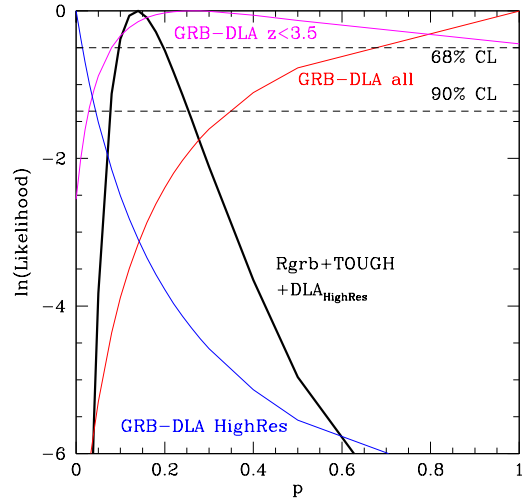}
\end{center}
\caption{{\rev Likelihood for the value of the metal-independent
    channel for GRB production ($p$), derived from analysis of the
    metallicity of GRB-DLAs, using data from
    \citet{cucchiara2014}. The red line shows the result from analysis
    of the full sample; the magenta line for the subsample of $z<3.5$
    data, while the blue line is for the subsample of high-resolution
    data. The black solid line shows the combined likelihood when
    adding comoving GRB rate and TOUGH star formation rate constraints
    to the subsample of high-resolution data. Dashed horizontal lines
    denotes the likelihood boundaries at 68\% and 90\%
    confidence. Compared to Figure~\ref{fig:likelihood}, which
    included both high and low resolution data at $z<3.5$, the
    combined likelihood contours obtained using high-resolution data
    (black line) are slightly shifted toward smaller
    $p$.}}\label{fig:likelihood_appendix}
\end{figure}

In this paper we have used a minimal but successful model for the
redshift evolution of the luminosity function of star forming (Lyman
Break) galaxies to investigate production of long-duration GRBs and
properties of their host galaxies. We followed the framework
introduced in our earlier works on the connection between GRBs and
LBGs as complementary probes of star formation across cosmic time
\citep{trenti12,trenti2013}. Here, we have made predictions for the
luminosity, {\rev stellar mass functions, and metallicity
  distribution} of GRB hosts, depending on the presence and strength
of the GRB metallicity bias. The key findings of our modeling,
introduced in Sections~\ref{sec:model}-\ref{sec:results} and discussed
in Section~\ref{sec:tough}, are the following:
\begin{itemize}
\item The luminosity function of GRB hosts is connected to that of
  star forming galaxies, but changes in the shape are introduced by
  several factors.  First, GRBs trace star formation, therefore to
  first order UV-bright hosts are preferred, with an approximate
  scaling given by host luminosity ($\Phi_{(GRB)}\propto
  \Phi_{(LBG)}\times L$). The presence of a metallicity bias
  qualitatively counteracts, at least partially, this luminosity
  function flattening, making GRBs more likely to have faint
  hosts. This established picture is however missing one key
  ingredient, which we discussed and modeled, namely the impact of
  dust reddening. Since dust is proportional to metallicity, and dust
  masks star formation, we highlight that to first approximation the
  dust has an impact on the GRB host luminosity function that is
  comparable but opposite to that of a mild metallicity bias (Figure~\ref{fig:metal_dust_balance}). Analysis
  of GRB host observations need to take that into account properly or
  else they would risk to draw (partially) incorrect information on
  the strength of the GRB metallicity bias.

\item {\rev The maximum likelihood model derived from analysis of the
    observed comoving GRB rate} is one that has a moderate metallicity
  bias, with about 80\% of GRBs in a very low metallicity environment
  produced by collapsars and the remaining 20\% by a metal-independent
  channel {\rev (such as, e.g. binaries or magnetar engines; see also \citealt{jimenez13})}. However, this
  model predicts that the large majority of $z\lesssim 1$ GRBs are
  produced by the metal independent channel, because low-$z$ hosts
  have metallicities where there is a preferential suppression of
  collapsars (Figure~\ref{fig:channel}). At intermediate redshifts
  ($z\sim 3$) the two channels produce comparable numbers of bursts,
  offering the most discriminating power between alternative
  scenarios. Finally, at $z\gtrsim 5$ most of the GRBs are produced in
  metal poor environments where collapsar efficiency has plateaued, so
  it becomes again hard to discriminate among models with different
  $p$.

\item We make detailed predictions for the luminosity and stellar mass
  functions of GRB hosts, as illustrated in the key
  Figures~\ref{fig:lf_distribution}-\ref{fig:stellar_masses}. Because
    of the direct relation between UV luminosity and GRB production,
    a comparison with luminosity/star formation rates of host galaxies
    is recommended over the use of stellar masses which are only
    indirectly, and approximatively, linked to recent star formation. To avoid
    introducing uncontrolled systematic errors (such as preferential
    reporting of detections over null results), it is also fundamental
    to use only complete samples of GRB hosts free of selection effects. 

  \item As a first comparison with observations, we present a maximum
    likelihood analysis of the star formation rates of the complete
    sub-sample of TOUGH observations of GRB hosts by
    \citet{michal2012}. The sample size is small (11 hosts) and at low
    redshift, but nevertheless it provides a preliminary
    characterization of the metallicity bias from host studies using
    our framework. The inferred strength of the metallicity bias is
    {\rev lower ($p=0.04$) but} consistent with that of our canonical
    model within uncertainties, leading to a combined constraint
    $0.1<p<0.35$ at 90\% confidence. This means that both the strong
    metal bias ($p=0$) and the no bias scenarios ($p=+\infty$) are
    clearly ruled out.

  \item Our model predicts the metallicity distribution of GRB hosts
    as well. This is shown in
    Figure~\ref{fig:metallicity_distribution}, which highlights that
    characterizing the median (and the lower 50\%) of the metallicity
    distribution of GRB hosts seems to be a powerful indicator of the
    strength of the metallicity bias. In contrast, the top of the
    metallicity of GRB hosts is remarkably similar among different
    models, providing relatively little insight to constrain $p$. The
    predictions shown in Figure~\ref{fig:metallicity_distribution} for
    $p=0.2$ appear in conflict with the {\rev low-redshift
      observations of GRB host metallicities by
      \citet{savaglio09}}. This suggests that smaller $p$ values
    smaller or closer to the peak of the likelihood from GRB host
    luminosities at $p=0.04$ yield an overall better description of
    the observations. However, we can make this statement only in a
    qualitative sense, since we do not have a complete sample of GRB
    host metallicities, {\rev and observational effects can strongly
      affect a formal  maximum likelihood analysis, as shown in
      Figure~\ref{fig:likelihood_appendix}.}

\end{itemize}

These conclusions allow us to address the two questions we posed in
Section~\ref{sec:intro} as one of the motivations to our study: Are
GRBs more probable in low-metallicity environments compared to higher
metallicity? Is there a maximum metallicity cut-off?  We conclude that
there is clear evidence for a metal-dependent relation between GRB and
star-formation rate, with low metallicity environments preferentially
producing bursts (see also \citealt{jimenez13}). However, a sharp
metallicity cut-off is strongly ruled out by the data-model
comparison. At the current time, the main limitation of the analysis
we presented is given by the lack of a well defined complete sample of
GRB host observations at intermediate redshifts. This should be the
top priority for further progress in the characterization of the
physics of GRB explosions from studies of their host galaxies.

{\rev Because our canonical model parameter $p=0.2$, inferred
  primarily from the GRB rate modeling, appears in tension with
  metallicity measurements, it is important to focus on acquiring more
  data on star formation rates and metallicities of GRB hosts.} Future
observations should also be able to test directly model consistency,
namely the three main ingredients we used to construct predictions:
(1) calibration of the luminosity functions starting from LBG
observations; (2) empirical, observationally motivated relations for
metallicity versus mass/luminosity (with its extrapolation to fainter
than observed galaxies) and for dust content of GRB hosts; (3) GRB
efficiency versus metallicity described by a simple relation with one
free parameter ($p$). For example, ALMA observations of GRB hosts have
the potential to characterize star formation rates, and measure
directly dust and metal content from molecular line diagnostic,
bypassing the dust absorption modeling present in the current work
focused on rest-frame UV data. Finally, in a few years, 30m class
observatories from the ground, and the \emph {James Webb Space
  Telescope}, will not only be capable of detecting fainter hosts than
those seen with current facilities, but also provide spectra of the
hosts seen today, thereby measuring directly the metallicity
distribution of the environments in which GRBs explode.

\acknowledgements 

We thank Sandra Savaglio for useful discussions, Nino Cucchiara for
sharing data,  and an anonymous
  referee for helpful comments. This work was partially supported by
the European Commission through the Marie Curie Career Integration
Fellowship PCIG12-GA-2012-333749 (MT) and by NSF Grant No. AST 1009396
(RP). RJ acknowledges support from Mineco grant
FPA2011-29678-C02-02. {\rev Tabulated model predictions are available in
electronic format.}



\begin{deluxetable}{rrrrrrrrr}
\tabletypesize{\footnotesize}
\tablecolumns{9}
\tablewidth{0pt}
\tablecaption{Schechter fits for GRB-host LF (rest-frame UV)) \label{tab:LFhost}}
\tablehead{ \colhead{$z$}  & \multicolumn{2}{c}{$p=0$} &
  \multicolumn{2}{c}{$p=0.04$}  & \multicolumn{2}{c}{$p=0.20$} &
   \multicolumn{2}{c}{NO BIAS} \\ \colhead{}  & \colhead{$M_{AB}^{(*)}$}  &
  \colhead{$\alpha$} &  \colhead{$M_{AB}^{(*)}$}  &
  \colhead{$\alpha$}  & \colhead{$M_{AB}^{(*)}$}  & \colhead{$\alpha$} & \colhead{$M_{AB}^{(*)}$}  & \colhead{$\alpha$}
  }
\startdata
0.25 & -20.3 & -1.2 & -19.4 & -0.1 & -19.4 & -0.1 & -19.4 & -0.1\\
0.75 & -20.2 & -1.4 & -19.8 & -0.3 & -19.7 & -0.1 &-19.7 &-0.1 \\
1.25 & -20.1 & -1.3 & -20.1 & -0.5 &-20.0 & -0.2 & -20.0 & -0.1  \\
1.75 & -20.2 & -1.1 & -20.3 & -0.6 &-20.2 & -0.3 & -20.2 & -0.1 \\
2.25 & -20.3 & -0.9 & -20.5 & -0.7 &-20.5 & -0.4 & -20.4 & -0.2 \\
2.75 & -20.7 & -0.8 & -20.9 & -0.8 &  -20.9 & -0.6 & -20.9 & -0.4 \\
3.25 & -20.9 & -0.8 & -21.0 & -0.8 & -21.0 & -0.7 & -21.0 & - 0.5\\
3.75 & -21.0 & -0.9 & -21.1 & -0.8 &-21.1 & -0.7 & -21.1 & -0.5 \\
4.25 & -21.0 & -0.9 & -21.1 & -0.9 & -21.2 & -0.8 & -21.1 & -0.6  \\
4.75 &-21.1 & -1.0 & -21.2 & -0.9 & -21.3 & -0.8 & -21.2 & -0.6 \\
5.25 & -21.3 & -1.0 & -21.3 & -1.0 &-21.4 & -0.9 & -21.4 & -0.7\\
5.75 & -21.3 & -1.1 & -21.3 & -1.1 &-21.4 & -1.0 & -21.4 & -0.8 \\
6.00 &-21.3 & -1.1 & -21.3 & -1.1 &-21.4 & -1.0 & -21.4 & -0.8  \\
7.00 & -21.3 & -1.2 & -21.3 & -1.2 & -21.4 & -1.1 & -21.3 & -0.9 \\
8.00 & -21.2 & -1.3 & -21.2 & -1.3 & -21.3 & -1.2 & -21.3 & -1.0 \\
9.00 & -21.1 & -1.5 & -21.2 & -1.4 & -21.2 & -1.4 & -21.2 & -1.2 \\
\enddata
\tablenotetext{a}{The Schechter fit parameters provided above {\rev
    for models with different $p$ value} are valid in the interval
  $-22<M_{AB}<-17.5$. At fainter magnitudes the GRB host LF may
  exhibit significant differences, especially at low redshift where it
  becomes steeper.}
\end{deluxetable}


\begin{deluxetable}{rrrrrrrr}


\tabletypesize{\scriptsize}


\tablecaption{Model predictions for GRB host properties\label{tab:modeloutput}}

\tablenum{2}

\tablehead{\colhead{z} & \colhead{p} & \colhead{$\mathrm{Mag^{(w/dust)}_{UV}}$} &
  \colhead{$\mathrm{Mag_{UV}}$} & \colhead{$M_{*}$} & \colhead{$M_{DM}$} & \colhead{$\mathrm{\log_{10}(Z/Z_{\sun})}$} & \colhead{\#} \\
\colhead{} & \colhead{} & \colhead{} & \colhead{} & \colhead{} & \colhead{} & \colhead{} & \colhead{} }

\startdata
3.75 & 0.2 & -2.391e+01 & -2.633e+01 & 4.566e+11 & 9.682e+13 & -1.462e-02 & 2.564e-08 \\
3.75 & 0.2 & -2.389e+01 & -2.630e+01 & 4.451e+11 & 9.527e+13 & -1.673e-02 & 3.996e-08 \\
3.75 & 0.2 & -2.386e+01 & -2.626e+01 & 4.301e+11 & 9.375e+13 & -1.962e-02 & 6.403e-08 \\
3.75 & 0.2 & -2.384e+01 & -2.623e+01 & 4.193e+11 & 9.224e+13 & -2.177e-02 & 8.609e-08 \\
3.75 & 0.2 & -2.382e+01 & -2.620e+01 & 4.106e+11 & 9.077e+13 & -2.358e-02 & 1.105e-07 \\
\enddata


\tablecomments{Table 2 is published in its entirety in the electronic
edition of the {\it Astrophysical Journal}.  A portion is shown here for guidance regarding its form and content.}


\end{deluxetable}

\end{document}